\documentclass{iopart}

\usepackage{iopams,amsfonts,amssymb,amsthm} 
\usepackage{geometry}                
\usepackage{graphicx}
\usepackage{rawfonts}
\usepackage{amssymb}
\usepackage{epstopdf}
\usepackage[usenames,dvipsnames,svgnames]{xcolor}
\usepackage{color}
\usepackage[colorlinks=true,citecolor=blue,linkcolor=red,urlcolor=blue]{hyperref}
\usepackage{comment}

\input prepictex.tex
\input pictex.tex
\input postpictex.tex

\newtheorem{theo}{Theorem}
\newtheorem{lemm}{Lemma}

\def\Pr{\noindent \emph{Proof: }}
\def\qed{$\Box$}

\def\nor{\normalsize}

\def\sfrac#1#2{\hbox{\nor $\frac{#1}{#2}$}}
\def\Sfrac#1#2{\hbox{\large $\frac{#1}{#2}$}}

\def\BRef#1{(\ref{#1})}

\begin{document}

\title[Copolymeric Combs]{Force-induced desorption
of copolymeric comb polymers}

\author{E J Janse van Rensburg$^\star$, C E Soteros$\ddagger$ \& S G Whittington$\dagger$}

\address{
$^\star$ Department of Mathematics and Statistics, York University, Toronto,  M3J 1P3, Canada\\ 
$\ddagger$ Department of Mathematics and Statistics,  University of Saskatchewan, Saskatoon,  S7N 5E6, Canada\\ 
$\dagger$ Department of Chemistry,  University of Toronto, Toronto,  M5S 3H6, Canada}
\ead{
$^\star$rensburg@yorku.ca, 
$^\ddagger$soteros@math.usask.ca,
$^\dagger$stuart.whittington@utoronto.ca}
\vspace{10pt}
\begin{indented}
\item[] \today
\end{indented}

\begin{abstract}
We investigate a lattice model of comb copolymers that can adsorb at a surface and 
that are subject to a force causing desorption.  The teeth and the backbone of the 
comb are chemically distinct and can interact differently with the surface.  
That is, the strength of the surface interaction can be different for the monomers 
in the teeth and in the backbone.  We consider several cases including (i) the 
uniform case where the number of teeth is fixed and the lengths of the branches 
in the backbone and the lengths of the teeth are all identical, (ii) the case where 
the teeth are short compared to the branches in the backbone, (iii) the situation 
where the teeth are long compared to the backbone, and (iv) the case where the 
number of teeth approaches infinity.  We determine the free energies in the 
thermodynamic limit and discuss the nature of the phase diagrams of the model.

\end{abstract}

\pacs{82.35.Lr,82.35.Gh,61.25.Hq}
\ams{82B41, 82B80, 65C05}

\vspace{2pc}
\noindent{\it Keywords}: Self-avoiding walk, copolymer combs, phase diagram

%
\maketitle
%
%

\section{Introduction}
\label{sec:Introduction}

The introduction of experimental techniques such as atomic force microscopy 
\cite{Haupt1999,Zhang2003} has made possible the micro-manipulation of 
individual polymer molecules and allowed the investigation of how individual 
polymer molecules respond to applied forces.  This has led to renewed 
interest in the theoretical treatment of this topic
\cite{Beaton2015,IoffeVelenik,IoffeVelenik2010,Rensburg2016a}.
For a review, see for instance \cite{Orlandini}.

A particular case that has attracted attention is when the polymer is adsorbed 
at a surface and is desorbed by the action of the force 
\cite{Guttmann2014,Rensburg2013,Rensburg2017,Krawczyk2005,Krawczyk2004,Mishra2005}.
Several different models have been investigated \cite{Orlandini} but 
we shall concentrate here on interacting models of self-avoiding walks
\cite{Rensburg2015,MadrasSlade} and related systems.  For other related work, 
see \cite{Beaton2017,Skvortsov2009,Binder2012}.

\begin{figure}[h!]
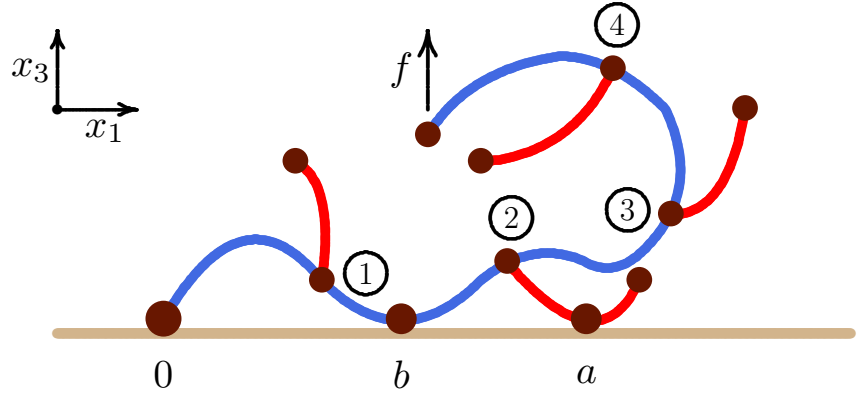

\beginpicture
\setcoordinatesystem units <1pt,1pt>
\setplotarea x from -70 to 200, y from -20 to 110

\color{Tan}
\setplotsymbol ({\scalebox{1.0}{$\bullet$}})
\plot 0 -5 300 -5 /

\color{RoyalBlue}
\setplotsymbol ({\scalebox{0.9}{$\bullet$}})
\setquadratic
\plot 40 0 70 30 100 15 130 0 160 15 180 25 200 22 210 20 220 25 235 50 
230 80 210 95 190 100 160 90 140 70 /

\color{Red}
\plot 100 15 100 45 90 60 /
\plot 170 22 200 0 220 15 /
\plot 232 40 250 50 260 80 /
\plot 210 95 190 70 160 60 /

\color{Sepia}
\multiput {\scalebox{2.5}{$\bullet$}} at 100 15 170 22 232 40 210 95 140 70
90 60 220 15 260 80 160 60 /
\multiput {\scalebox{3.5}{$\bullet$}} at 40 0 /
\multiput {\scalebox{3.0}{$\bullet$}} at 130 0 200 0 /

\color{black}
\setplotsymbol ({\scalebox{0.33}{$\bullet$}})
\arrow <10pt> [.2,.67] from 140 80 to 140 110 
\put {\scalebox{1.5}{$f$}} at 130 95
\put {\scalebox{1.5}{$0$}} at 40 -20
\put {\scalebox{1.5}{$b$}} at 130 -20
\put {\scalebox{1.5}{$a$}} at 200 -20

\arrow <7pt>  [.2,.67] from 0 80 to 30 80 \put {$\scalebox{1.5}{$x_1$}$} at 18 72
\arrow <7pt>  [.2,.67] from 0 80 to  0 110 \put {$\scalebox{1.5}{$x_3$}$} at -10 95
\put {$\bullet$} at 0 80

\put {
 \beginpicture
\circulararc 360 degrees from 8 0 center at 0 0 
 \put {\large$1$} at 0 0 
 \endpicture
 } at 115 18

\put {
 \beginpicture
\circulararc 360 degrees from 8 0 center at 0 0 
 \put {\large$2$} at 0 0 
 \endpicture
 } at 170 39

\put {
 \beginpicture
\circulararc 360 degrees from 8 0 center at 0 0 
 \put {\large$3$} at 0 0 
 \endpicture
 } at 214 42

\put {
 \beginpicture
\circulararc 360 degrees from 8 0 center at 0 0 
 \put {\large$4$} at 0 0 
 \endpicture
 } at 210 112

\color{black}
\normalcolor
\endpicture
\caption{A schematic diagram of a pulled adsorbing comb block copolymer.  
The comb consists of a backbone (blue) consisting of $B$-type monomers 
and $t$ teeth (red) consisting of $A$-type monomers. An endpoint of 
the comb's backbone  is grafted to the origin on the adsorbing 
surface $x_3=0$, and the other endpoint of its backbone is 
pulled vertically by a force $f$. Monomers along the backbone adsorb on 
the adsorbing surface with activity $b = e^{-\epsilon_B/k_BT}$ where 
$\epsilon_B$ is a binding energy, $T$ is the absolute temperature
and $k_B$ is Boltzmann's constant.  Monomers in the teeth of the comb
similarly adsorb on the adsorbing surface, but with activity 
$a=e^{-\epsilon_A/k_BT}$ where $\epsilon_A$ is a binding energy.  
Starting from the origin and moving along the backbone, the trivalent 
vertices (branch points) are labelled $1,2, ...,t$ as indicated in 
circles, where the number of teeth $t=4$ in this case. The teeth are 
similarly labelled according to the label of the trivalent vertex 
from which they emanate. Branches of the backbone are labelled 
with $1,...,t+1$ so that the branch between the trivalent vertices 
$i-1$ and $i$ has label $i$. We assume that each backbone branch has 
length $m_b$ and each of the teeth has length $m_a$. }
\label{F1oo}
\end{figure}

Polymer adsorption is important in steric stabilization of colloids 
\cite{Napper} and linear polymers terminally attached to a surface can 
be used as steric stabilizers.  Other polymer architectures such as stars 
and brushes have been used for this purpose.  Block copolymers are also 
useful for steric stabilization, where one block adsorbs on 
the surface and the other extends into the solution.  In particular, 
comb polymers with the backbone and teeth having different chemical 
composition (see figure \ref{F1oo}) have been investigated 
experimentally \cite{Xie}.  These have several parameters that can 
be varied, including the length and spacing of the teeth of the comb 
and the chemical compositions of the teeth and backbone.

There are a few studies of the self-avoiding walk model of 
homopolymeric star polymers, adsorbed at a surface and subject to 
a desorbing force \cite{Bradly,BradlyOwczarek,Rensburg2018,Rensburg2019}.
For an investigation of the same situation for some models of 
homopolymeric branched polymers, including uniform combs, see
reference \cite{Rensburg2019}.  Even fewer papers 
\cite{RensburgSoterosWhittington,Rensburg2022} have appeared on 
self-avoiding walk models of copolymers adsorbed at a surface and subject 
to a force.  Linear $AB$ diblock copolymers and $ABA$ triblock copolymers 
have been studied where the $A$ and $B$ monomers interact differently with 
the surface \cite{RensburgSoterosWhittington}.  One end monomer, \emph{ie} 
a vertex of degree 1, of the copolymer is grafted to the surface and the force 
is applied either at the opposite end (the other degree 1 vertex) of 
the block copolymer or at the central vertex.  Copolymeric stars have 
also been considered \cite{Rensburg2022}.  Think of a multi-arm star with some 
arms (or branches) of type $A$ and the remaining arms of type $B$.  A vertex of 
degree 1 is fixed in the surface (say of an $A$-arm) and the force is 
applied either at a degree 1 vertex of type $A$ or $B$, or at the central 
vertex.  In each case the general form of the phase diagram 
has been established \cite{Rensburg2022}.

In this paper we explore the problem of copolymeric comb polymers where 
the backbone and the teeth have different chemical composition; see 
figure \ref{F1oo} for a schematic diagram.  This is a generalization 
and extension of a treatment of a self-avoiding walk model of 
homopolymeric combs, pulled from a surface at which they adsorb, studied 
in \cite{Rensburg2019}.  There it was shown that there is a critical force 
where the comb is pulled from an adsorbed phase into a ballistic phase.  
An applied force at the terminal end of the backbone may be strong enough 
to pull off both the last segment of the backbone and the last tooth.  In 
this case it desorbs the entire comb.  There is also the possibility that 
the force is strong enough to pull off the last segment (or branch) of the 
backbone, but not the last tooth.  In this case the teeth remain adsorbed, 
and the comb only desorbs completely if the strength of the force is 
increased to also pull off the last tooth.

\begin{figure}[h!]
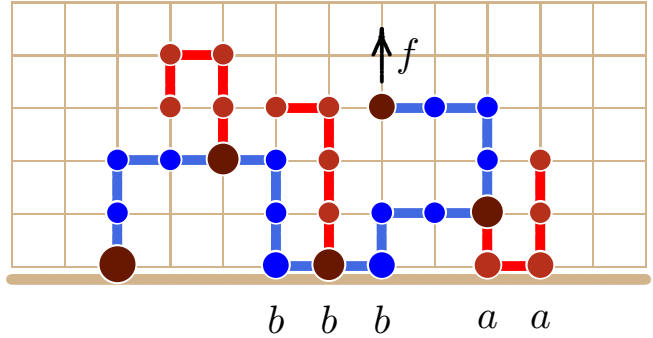

\beginpicture
\setcoordinatesystem units <1pt,1pt>
\setplotarea x from -70 to 240, y from -20 to 110

\color{Tan}
\setplotarea x from 0 to 240, y from 0 to 100
\setplotsymbol ({\scalebox{0.1}{$\bullet$}})
\grid 12 5
\setplotsymbol ({\scalebox{1.0}{$\bullet$}})
\plot 0 -5 240 -5 /

\setplotsymbol ({\scalebox{0.9}{$\bullet$}})
\color{RoyalBlue}
\plot 40 0 40 40 80 40 100 40 100 0 140 0 140 20 160 20
180 20 180 40 180 60 160 60 140 60 /
\color{White}
\multiput{\scalebox{2.4}{$\bullet$}} at 40 20 40 40 60 40 80 40 100 40 
100 20 140 20 160 20 180 20 180 40 180 60 160 60 140 60 /
\multiput{\scalebox{2.9}{$\bullet$}} at 100 0 120 0 140 0 /
\color{Blue}
\multiput{\scalebox{2.0}{$\bullet$}} at 40 20 40 40 60 40 80 40 100 40 
100 20 140 20 160 20 180 20 180 40 180 60 160 60 140 60 /
\multiput{\scalebox{2.5}{$\bullet$}} at 100 0 120 0 140 0 /

\setplotsymbol ({\scalebox{0.9}{$\bullet$}})
\color{Red}
\plot 80 40 80 80 60 80 60 60 /
\plot 120 0 120 60 100 60 /
\plot 180 20 180 0 200 0 200 20 200 40 /
\color{White}
\multiput{\scalebox{2.4}{$\bullet$}} at 80 40 80 60 80 80 60 80 60 60
120 0 120 20 120 40 120 60 100 60 180 20 180 0 200 0 200 20 200 40 /
\multiput{\scalebox{2.9}{$\bullet$}} at 180 0 200 0 /
\color{BrickRed}
\multiput{\scalebox{2.0}{$\bullet$}} at 80 40 80 60 80 80 60 80 60 60
120 0 120 20 120 40 120 60 100 60 180 20 180 0 200 0 200 20 200 40 /
\multiput{\scalebox{2.5}{$\bullet$}} at 180 0 200 0 /

\color{White}
\multiput {\scalebox{2.9}{$\bullet$}} at 140 60  /
\multiput {\scalebox{3.9}{$\bullet$}} at 40 0 /
\multiput {\scalebox{3.4}{$\bullet$}} at 80 40 120 0 180 20 /
\color{Sepia}
\multiput {\scalebox{2.5}{$\bullet$}} at 140 60  /
\multiput {\scalebox{3.5}{$\bullet$}} at 40 0 /
\multiput {\scalebox{3.0}{$\bullet$}} at 80 40 120 0 180 20 /

\color{black}
\setplotsymbol ({\scalebox{0.33}{$\bullet$}})
\arrow <10pt> [.2,.67] from 140 70 to 140 90 
\put {\scalebox{1.5}{$f$}} at 150 80
\multiput {\scalebox{1.5}{$a$}} at 180 -20 200 -20 /
\multiput {\scalebox{1.5}{$b$}} at 100 -20 120 -20 140 -20 /

\color{black}
\normalcolor
\endpicture
\caption{A lattice copolymeric comb.  The backbone of the comb adsorbs with
activity $b$, and the teeth with activity $a$.  A vertical force
$f$ pulls the comb at its endpoint, while its other endpoint is grafted
at the origin.  This comb has $3$ teeth.}
\label{F2}
\end{figure}

A lattice model of a copolymeric comb is illustrated in figure \ref{F2}.
The comb is embedded in the cubic lattice ${\mathbb Z}^3$ (or more
generically, in the $d$-dimensional hypercubic lattice).  Attaching
the coordinates $(x_1,x_2,x_3)$ to the lattice sites (with each coordinate
an integer), the comb in figure \ref{F2} is also in the \textit{half-lattice}
${\mathbb L}^3$ of all lattice sites with $x_3\geq 0$.  The half-lattice
has a boundary (hard wall) $x_3=0$ that will also be called the
\textit{adsorbing surface} or \textit{adsorbing plane}.

The backbone of the comb in figure \ref{F2} is composed of $B$-monomers, 
and has length $16$ consisting of $4$ segments or \textit{branches}, 
each of length $m_b=4$.  The comb has $t=3$ teeth composed of 
$A$-monomers, each of length $m_a=3$.  The total size of the comb 
is $(t+1)\,m_b+t\,m_a = 28$.  The backbone of 
the comb makes $v_B=3$ \textit{visits}, each of weight $b$, to the adsorbing 
plane, and the teeth make $v_A=2$ visits, each of weight $a$.  The last 
vertex of the backbone is at a height $h=3$ above the adsorbing plane.  
Putting $y=e^{f/k_BT}$, where $f$ is the applied force, $k_B$ is 
Boltzmann's constant and $T$ is the absolute temperature, the partition 
function of the comb is 
\begin{equation}
\label{eq:genpf}
 K^{(t)}(m_a,m_b,a,b,y) = \sum_{v_A,v_B,h} k^{(t)}(m_a,m_b,v_A,v_B,h)\,
 a^{v_A}b^{v_B}y^h ,
\end{equation}
where $k^{(t)}(m_a,m_b,v_A,v_B,h)$ is the number of combs, with 
one terminal vertex attached to the surface, with $t$ teeth, each 
of length $m_a$, $t+1$ backbone branches of length $m_b$, having 
$v_A$ tooth vertices in the surface and $v_B + 1$ backbone vertices 
in the surface,  and with the other terminal vertex at height $h$ 
above the surface.  The thermodynamic limit of this model can be 
examined in several cases.  In the first instance, putting 
$m_b=m_a=m$ gives a uniform comb, and taking $m\to\infty$ 
(with $t$ fixed) gives a sparsely branched comb with a long backbone 
and long teeth. Other limits can be taken by taking $m_b\to\infty$ 
with $m_a$ fixed, or taking $m_a\to\infty$ with $m_b$ fixed.  There 
is also the limit $t\to\infty$, followed by either $m_b\to\infty$ 
or $m_a\to\infty$, or both.  We shall investigate some of these limits.

The plan of the paper is as follows.  In section \ref{sec:review} we 
recall some results about self-avoiding walks adsorbing at a surface 
and subject to a desorbing force, and in the same section we prove some 
lemmas that we shall need later in the paper.

In section \ref{sec:combs} we look at the case of uniform combs where 
$m_a=m_b=m$ (so that all branches in the backbone and teeth are the 
same length) and the number of teeth, $t$, is fixed.  We derive expressions 
for the free energy and give the general features of the phase diagram as 
slices at constant $a$, constant $b$ and constant $y$.  In 
section \ref{sec:nonuniform} we examine the situations where the 
backbone is long compared to the teeth, and where the teeth are 
long compared to the backbone, still with $t$ fixed.  These cases 
give rise to two limiting cases, one where the comb behaves like a 
self-avoiding walk and the other where it behaves like a star. In 
section \ref{sec:infinitet} we examine the infinite $t$ limit, and
we close with a short discussion in section \ref{sec:discussion}.

\section{Some background results}
\label{sec:review}

\subsection{Adsorbing and pulled walks}
\label{sec:2.1}
In this section we give a brief account of previous results, concentrating 
on self-avoiding walks.  We shall need some of these results in the following 
sections where we construct combs from collections of self-avoiding walks 
and stars.  We focus on the simple cubic lattice $\mathbb{Z}^3$ but some 
of the results for self-avoiding walks can be extended to $\mathbb{Z}^d$ 
for all $d \ge 2$.  

Consider self-avoiding walks in $\mathbb{Z}^3$ where we attach the standard
coordinate system $(x_1,x_2,x_3)$ so that lattice vertices have integer 
coordinates. For an $n$-edge self-avoiding walk, number the vertices 
$k=0,1, \ldots, n$ and write $x_i(k)$ for the $i$-th coordinate of the 
$k$-th vertex.  Write $c_n$ for the number of $n$-edge self-avoiding 
walks starting at the origin.  It is known that the limit
\begin{equation}
\lim_{n\to\infty} \Sfrac{1}{n}  \log c_n = \inf_{n>0} 
\Sfrac{1}{n}  \log c_n \equiv \log \mu_3
\end{equation}
exists \cite{Hammersley1957} and $\mu_3$ is called the 
\emph{growth constant} of the lattice (while $\kappa_3=\log \mu_3$ is the 
\emph{connective constant} of the lattice).

An $n$-edge self-avoiding walk is \emph{unfolded in the $x_i$-direction} 
(or \emph{$x_i$-unfolded}) if $x_i(0) < x_i(k) \le x_i(n)$ for all 
$1\le k \le n$.  It has \emph{$x_i$-span} equal to $x_i(n)-x_i(0)$.  
We shall write $c_n^{\dagger}$ (since the count is independent of $i$) 
for the number of these unfolded walks.  It is known \cite{HammersleyWelsh} 
that
\begin{equation}
c_n^{\dagger} = \mu_3^{n + O(\sqrt{n})}.
\label{eqn:HammersleyWelsh}
\end{equation}
In general we use the superscript $\dagger$ to show that a walk is unfolded 
and $\dagger [i]$ when we need to emphasise that the unfolding is in the 
$i$-th direction.

If the walk is \emph{doubly unfolded}, then it is unfolded in one direction 
and then unfolded again in a second direction, and we use the superscript 
$\ddagger$, or $\ddagger [ij]$ if we need to specify the two unfolding 
directions.

\begin{figure}[h!]
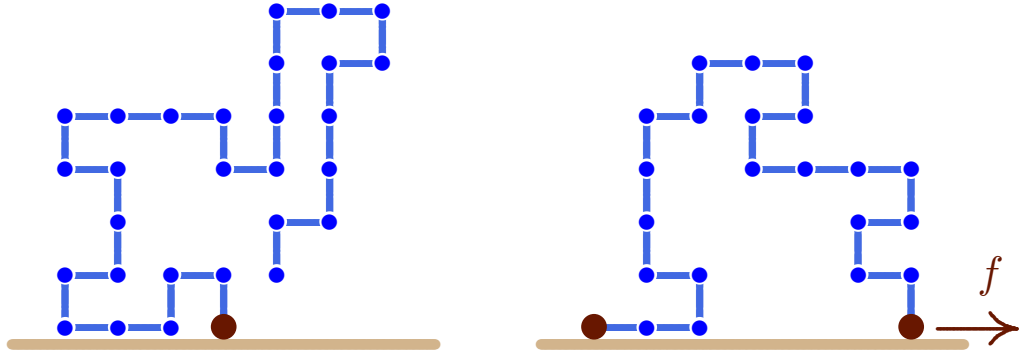

\beginpicture
\setcoordinatesystem units <2pt,2pt>
\setplotarea x from -20 to 250, y from -10 to 65
\setplotsymbol ({$\bullet$})
\color{Tan} \plot 0 -3  80 -3 /
\setplotsymbol ({\scalebox{0.66}{$\bullet$}})
\color{RoyalBlue}
\plot 40 0 40 10 30 10 30 0 20 0 10 0 10 10 20 10 20 20 20 30 10 30 10 40 20 40 
30 40 40 40 40 30 50 30 50 40 50 50 50 60 60 60 70 60 70 50 60 50 60 40 60 30 
60 20 50 20 50 10 /
\color{white}
\multiput {\scalebox{2.0}{$\bullet$}} at
40 0 40 10 30 10 30 0 20 0 10 0 10 10 20 10 20 20 20 30 10 30 10 40 20 40 
30 40 40 40 40 30 50 30 50 40 50 50 50 60 60 60 70 60 70 50 60 50 60 40 60 30 
60 20 50 20 50 10 /
\color{blue}
\multiput {\scalebox{1.5}{$\bullet$}} at
40 0 40 10 30 10 30 0 20 0 10 0 10 10 20 10 20 20 20 30 10 30 10 40 20 40 
30 40 40 40 40 30 50 30 50 40 50 50 50 60 60 60 70 60 70 50 60 50 60 40 60 30 
60 20 50 20 50 10 /
\color{Sepia} \put {\scalebox{2.5}{$\bullet$}} at 40 0 

\setcoordinatesystem units <2pt,2pt> point at -100 0 
\setplotsymbol ({$\bullet$})
\color{Tan} \plot 0 -3  80 -3 /
\setplotsymbol ({\scalebox{0.66}{$\bullet$}})
\color{RoyalBlue}
\plot 10 0 20 0 30 0 30 10 20 10 20 20 20 30 20 40 30 40 30 50 40 50 50 50 
50 40 40 40 40 30 50 30 60 30 70 30 70 20 60 20 60 10 70 10 70 0 /
\color{white}
\multiput {\scalebox{2.0}{$\bullet$}} at
10 0 20 0 30 0 30 10 20 10 20 20 20 30 20 40 30 40 30 50 40 50 50 50 
50 40 40 40 40 30 50 30 60 30 70 30 70 20 60 20 60 10 70 10 70 0 /
\color{blue}
\multiput {\scalebox{1.5}{$\bullet$}} at
10 0 20 0 30 0 30 10 20 10 20 20 20 30 20 40 30 40 30 50 40 50 50 50 
50 40 40 40 40 30 50 30 60 30 70 30 70 20 60 20 60 10 70 10 70 0 /
\color{Sepia} \multiput {\scalebox{2.5}{$\bullet$}} at 10 0 70 0 / 

\setplotsymbol ({\scalebox{0.33}{$\bullet$}})
\plot 75 0 90 0 /
\setquadratic
\plot 90 0 87 1 85 3 /
\plot 90 0 87 -1 85 -3 /
\put {\scalebox{1.67}{$f$}} at 85 10  

\color{black}
\normalcolor
\endpicture
\caption{(Left) A positive walk in the half-space $x_3\geq 0$.  (Right) An
$x_1$-unfolded loop in the half-space $x_3\geq 0$.  This loop is pulled at its
last vertex by a horizontal force $f$ parallel to the hard wall (boundary
of the half-space).}
\label{Fig1}
\end{figure}

If a self-avoiding walk satisfies the constraint that 
$x_3(k) \ge 0$ for $0 \le k \le n$ then the walk is a \emph{positive walk}  
(see figure \ref{Fig1}).  A positive walk can also be unfolded.  If 
we write $c_n^+$ for the number of positive walks with $n$ edges, and 
$c_n^{+\dagger [3]}$ for the number of positive walks with $n$ edges, 
unfolded in the $x_3$-direction, then 
$\lim_{n\to\infty} \sfrac{1}{n} \log c_n^+ =
\lim_{n\to\infty} \sfrac{1}{n} \log c_n^{+\dagger [3]} = \log \mu_3$
\cite{HammersleyWelsh,Whittington1975}.

If a positive walk is constrained to start and end in the plane $x_3=0$, then 
it is a \textit{loop}.  If we write $\ell_n$ for the number of loops 
with $n$ edges, and $\ell_n^{\dagger [1]}$ for the number of loops, 
unfolded in the $x_1$-direction, with $n$ edges, then 
$\lim_{n\to\infty} \sfrac{1}{n} \log \ell_n =
\lim_{n\to\infty} \sfrac{1}{n} \log \ell_n^{\dagger [1]} = \log \mu_3$
\cite{HammersleyWelsh,Whittington1975}. By symmetry, the same 
result holds for loops unfolded in the $x_2$-direction. 

All these results are also valid in the square lattice (or, more generally,
in the hypercubic lattices).

Let $c_n^+(v,h)$ be the number of $n$-edge positive walks, starting at 
the origin, having $v+1$ vertices in the plane $x_3=0$ and with $x_3(n)=h$.
We say that the walk has $v$ \emph{visits} and that the \emph{height} 
of the last vertex is $h$. Define the partition function 
$C_n^+(a,y) = \sum_{v,h} c_n^+(v,h)\, a^vy^h$.  The limit 
\begin{equation}
\lim_{n\to\infty} \Sfrac{1}{n}  \log \sum_{v,h} c_n^+(v,h)\, a^vy^h
=\lim_{n\to\infty} \Sfrac{1}{n}  \log C_n^+(a,y) \equiv \psi(a,y)
\label{eqn3}
\end{equation}
exists for all $a$ and $y$ \cite{Rensburg2013} and $\psi(a,y)$ is 
the \emph{free energy} of the model.  (We use lower case letters for 
counts, upper case letters for partition functions, and Greek letters 
for free energies.)  Here $a=\exp[-\epsilon/k_BT]$ 
and $y=\exp[f/k_BT]$ where $\epsilon$ is the energy associated with
a vertex in the surface, $k_B$ is Boltzmann's constant, $T$ is the 
absolute temperature and $f$ is the force normal to the surface 
(measured in energy units).

We can turn off the force by setting $y=1$ and we then have the pure 
adsorption problem where $\psi(a,1) \equiv \kappa(a)$.  The
free energy $\kappa(a)$ is a convex function of $\log a$ \cite{HTW} 
and is therefore continuous.  There exists a critical value of $a$, 
$a_c > 1$, such that $\kappa(a) = \log \mu_3$ when $a \le a_c$ 
and $\kappa(a) > \log \mu_3$ when $a > a_c$ \cite{HTW,Rensburg1998,Madras}. 

If we write $\ell_n(v)$ for the number of loops with $n$ edges and $v+1$ 
vertices in the plane $x_3=0$ then \cite{HTW} 
\begin{equation} 
\lim_{n\to\infty} \Sfrac{1}{n} \log \sum_v \ell_n(v)\, a^v 
 = \lim_{n\to\infty} \Sfrac{1}{n} \log L_n(a,1) = \kappa(a).
\end{equation}
We say that $\ell_n(v)$ is the number of adsorbing loops with $n$ 
edges and $v$ visits, and $L_n(a,1)$ is the partition function 
for loops with no applied force.  The free energy $\kappa(a)$ is 
unchanged when the walks or loops are unfolded, or even doubly 
unfolded \cite{HTW}, or when the endpoints of loops are restricted to 
a line (for example, to the $x_1$-axis), see, for example, reference 
\cite[section 9.1]{Rensburg2015}. Restricting any of
these walks or loops to a quarter space (for example $x_i\geq 0$
for $i\in\{1,2\}$) also leaves $\kappa(a)$ unchanged, even if
endpoints are restricted to be located in a  line.
For example, if adsorbing loops are unfolded in the $x_1$-direction 
(see figure \ref{Fig1}) their partition function is 
$L_n^{\dagger [1]}(a,1)$ and their free energy is also equal 
to $\kappa(a)$ \cite{HTW}.  The free energy is unchanged for 
the subset of these loops restricted to the quarter space
$x_2,x_3\geq 0$ and having end points in the $x_1$-axis.

We can turn off the interaction between a positive walk and 
the surface by setting $a=1$ so that the surface just acts as 
an impenetrable barrier (or hard wall) and we can then write 
$\psi(1,y) \equiv \lambda(y)$.  $\lambda(y)$ is a convex function 
of $\log y$ \cite{Rensburg2009}, $\lambda(y) = \log \mu_3$ for $ y \le 1$ 
and $\lambda(y) > \log \mu_3$ for $y > 1$ \cite{Beaton2015,IoffeVelenik}.
This is the free energy of pulled positive walks.  If the positive walks 
are unfolded, or doubly unfolded, in any direction, then the free energy
is unchanged.  For example, if the pulled walks are unfolded in the
$x_3$-direction, then the partition function is $C_n^{+\dagger[3]}(1,y)$
and the free energy is $\lambda(y)$ \cite{Rensburg2016a}.

Pulled positive walks are a special case of walks (not necessarily 
positive) terminally attached to the origin and then pulled at 
the other end by a force $f$ in the $x_3$ direction; we call 
these $x_3$-pulled walks. The partition function for these walks 
is denoted here by $C_n(y)$ with $y$ being conjugate to the 
$x_3$-span of the walk. $C_n(y)\geq C_n^{+}(1,y)$ and its 
limiting free energy is also $\lambda(y)=\psi(1,y)$ for $y\geq 1$.
This follows  from  unfolding arguments that give the string of inequalities 
\begin{equation}
C_n^+(1,y) \le C_n(y) 
\le C_n^{\dagger [3]}(y) e^{O(\sqrt{n})} 
\le C_n^+(1,y) e^{O(\sqrt{n})}\quad\hbox{for $y \ge 1$.}
\end{equation}

For general values of $a$ and $y$ \cite{Rensburg2013}
\begin{equation}
\psi(a,y) = \max[\kappa(a),\lambda(y)]
\label{eq:psidef}
\end{equation}
so $\psi(a,y) = \log \mu_3$ when $a \le a_c$ and $y \le 1$.  This is 
the \emph{free phase}.  There are phase boundaries in the $(a,y)$-plane 
at $a=a_c$ for $y \le 1$, at $y=1$ for $a \le a_c$ and at 
the solution of $\kappa(a) = \lambda(y) $ for $a \ge a_c$ and 
$y \ge 1$.  The phase diagram has three phases and the phase transition 
for $y > 1$ and $a > a_c$ between the \emph{adsorbed phase} and the 
\emph{ballistic phase} is first order \cite{Guttmann2014}.

The definitions for positive walks and loops above are all relative 
to a fixed boundary surface $x_3=0$ and a corresponding half-space 
$x_3\geq 0$. The same results hold for {\it negative} walks and loops 
in the half-space $x_3 \leq 0$.  Similarly, the results hold for 
other choices for the fixed boundary surface $x_i=0$, $i\in\{1,2\}$, 
and we define $x_i$-positive (negative) walks and loops for these 
cases. We also refer to $x_i$-loops as loops oriented in the $x_i$ 
direction.   Also by symmetry, results about $x_3$-pulled walks 
apply equally to $x_i$-pulled walks $i\in \{1,2\}$ where $y$ is 
conjugate to $x_i$-span. 

In some of the constructions that we use later in the paper we shall 
make extensive use of the properties of loops, and we shall need 
several new results about loops.  If the endpoint of a loop 
(oriented in the $x_3$-direction) is pulled by a force $f$ parallel 
to the $x_1$-direction (see the right panel of figure \ref{Fig1}), 
then it is both \textit{adsorbing and pulled}. We write $\ell_n(v,s)$ 
for the number of loops of $n$ edges with $v$ visits and $x_1$-span 
equal to $s=x_1(n)-x_1(0)$ (the distance between its endpoints in 
the $x_1$-direction), then, provided that the Boltzmann factor for 
adsorbed vertices is $a=1$ and $y\geq 1$,
\begin{equation}
\lim_{n\to\infty} \Sfrac{1}{n} 
\log L_n(1,y)
= \lim_{n\to\infty} \Sfrac{1}{n} 
\log \sum_{v,s} \ell_n (v,s)\, y^s =\lambda(y).
\end{equation}
This is also the case if $L_n(1,y)$ is replaced by the $x_i$-unfolded
loop partition function $L_n^{\dagger[i]}(1,y)$ for $i=1,2$, or
even for the partition function $L_n^{\ddagger[1,2]}(1,y)$ of
doubly unfolded loops in the $x_1$ and $x_2$ directions.  The
lemma also applies to these models when the walks are confined to a
quarter lattice defined by $x_2,x_3\geq 0$ and if the endpoints
of the loops are restricted to the $x_1$-axis and pulled parallel
to the $x_1$ axis.  We prove this result for the partition functions
$L_n^{\dagger[1]}(1,y)$ and $L_n(1,y)$ in the lemma \ref{lemm:lemma1}.  
The proofs for the other models are similar.
\begin{lemm}
\label{lemm:lemma1}
For loops with no interaction with the surface ($a=1$) and pulled 
parallel to the surface (in the $x_i$-direction, $i\in\{1,2\}$),  
the free energy is $\lambda(y)$ for $y\geq 1$. If the loops are 
unfolded in the $x_i$-direction they have free energy, $\lambda(y)$, 
for all $y>0$.
\end{lemm}
\Pr
Consider $i=1$ and $y\geq 1$.  Write $L_n(1,y)=\sum_{v,s} \ell_n (v,s)\, y^s$
for the partition function of loops with $n$ edges, pulled in the 
$x_1$-direction, and write $L_n^{\dagger [1]}(1,y)$ for the partition 
function of loops, unfolded and pulled in the $x_1$-direction.  Such 
loops are examples of $x_1$-pulled walks, thus by inclusion, for $y\geq 1$
$$L_n^{\dagger [1]}(1,y) \le L_n(1,y) \le C_n(y)= \exp[\psi(1,y) n + o(n)].$$

Consider self-avoiding walks unfolded in the $x_1$-direction and 
subject to a force in the same direction. These are examples of 
$x_1$-pulled $x_1$-positive walks.  Suppose that their partition 
function is $C_n^{\dagger [1]}(y)$.  Then $C_n^{\dagger [1]}(y)
\le C_n^{+[1]}(1,y)$ by inclusion.  Unfolding cannot decrease 
the distance between the end-points in the unfolding direction so, 
via equation (\ref{eqn:HammersleyWelsh}), for $y \ge 1$, 
$C_n^{+[1]}(1,y) \leq C_n(y) \leq C_n^{\dagger [1]}(y) e^{O(\sqrt{n})}$.
Therefore 
$$C_n^{\dagger [1]}(y) = \exp[\psi(1,y)n + o(n)].$$

Consider $n$-edge self-avoiding walks pulled and unfolded in the 
$x_1$-direction, and consider the subset of these walks unfolded 
in the $x_3$-direction. Write $C_n^{\ddagger [13]} (y)$ for their 
partition function.  Then 
$$C_n^{\ddagger [13]}(y) \le C_n^{\dagger [1]}(y) 
\le C_n^{\ddagger [13]}(y) \exp[O(\sqrt{n})].$$
Consider the subset of these walks with the {\it most popular span} 
in the $x_3$-direction.\footnote{The most popular $x_3$-span is that 
value of the span giving a maximal contribution to the partition function.  
That is, the most popular span refers to an $x_3$-span such that 
the contribution to the  partition function $C_n^{\ddagger [13]}(y)$ 
due to walks with that span is at least as large as the contribution 
due to any other $x_3$-span.}  By concatenating such a walk with
most popular span in the $x_3$ direction and the reverse of such 
a walk (joined by an additional edge) a loop of length $2n+1$ edges
is obtained, pulled and unfolded in the $x_1$-direction.  There are 
at least $\left(C_n^{\ddagger [13]}(y)/(n+1)\right)^2$ such loops 
obtained by this construction so
$$L_{2n+1}^{\dagger [1]}(1,y) 
\ge \left(\frac{C_n^{\ddagger [13]}(y)}{n+1}\right)^2 
 = \exp[ 2\, \psi(1,y) n + o(n)]$$
Therefore $L_n(1,y) = \exp[\psi(1,y) n + o(n)]$ and $L_n^{\dagger [1]}(1,y) 
= \exp[\psi(1,y) n + o(n)]$.

For $y<1$, the lemma is a corollary of lemma 1 and theorem 1 in 
reference \cite{Rensburg2016a}.
\qed

\begin{figure}[h!]
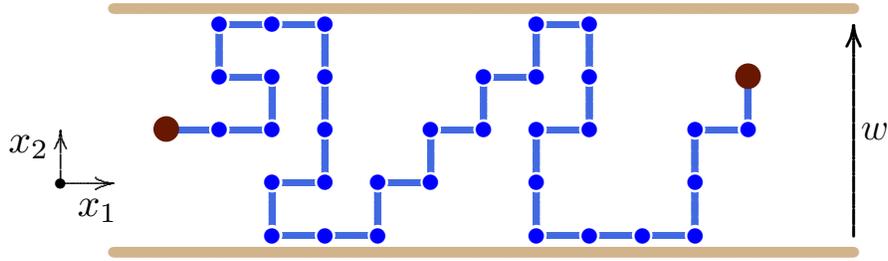

\beginpicture
\setcoordinatesystem units <2pt,2pt>
\setplotarea x from -35 to 200, y from -10 to 55

\color{black}
\arrow <7pt>  [.2,.67] from 0 10 to 10 10 
 \put {$\scalebox{1.5}{$x_1$}$} at 7 5
\arrow <7pt>  [.2,.67] from 0 10 to 0 20 
 \put {$\scalebox{1.5}{$x_2$}$} at -6 17
 \put {$\bullet$} at  0 10 

\setplotsymbol ({$\bullet$})
\color{Tan} \plot 10 -3  150 -3 /
\color{Tan} \plot 10 43  150 43 /
\setplotsymbol ({\scalebox{0.66}{$\bullet$}})
\color{RoyalBlue}
\plot 
20 20 30 20 40 20 40 30 30 30 30 40 40 40 50 40 50 30 50 20 50 10 
40 10 40 0 50 0 60 0 60 10 70 10 70 20 80 20 80 30 90 30 90 40 
100 40 100 30 100 20 90 20 90 10 90 0 100 0 110 0 120 0 120 10 
120 20 130 20 130 30 /
\color{white}
\multiput {\scalebox{2.0}{$\bullet$}} at
20 20 30 20 40 20 40 30 30 30 30 40 40 40 50 40 50 30 50 20 
50 10 40 10 40 0 50 0 60 0 60 10 70 10 70 20 80 20 80 30 90 30 
90 40 100 40 100 30 100 20 90 20 90 10 90 0 100 0 110 0 120 0 
120 10 120 20 130 20 130 30 /
\color{blue}
\multiput {\scalebox{1.5}{$\bullet$}} at
20 20 30 20 40 20 40 30 30 30 30 40 40 40 50 40 50 30 50 20 
50 10 40 10 40 0 50 0 60 0 60 10 70 10 70 20 80 20 80 30 90 30 
90 40 100 40 100 30 100 20 90 20 90 10 90 0 100 0 110 0 120 0 
120 10 120 20 130 20 130 30 /
\color{Sepia} \multiput {\scalebox{2.5}{$\bullet$}} at 20 20 130 30 / 

\color{black}
\setplotsymbol ({\scalebox{0.25}{$\bullet$}})
\arrow <8pt> [.2,.67] from 150 0 to 150 40
\put {\Large$w$} at 154 20 
\normalcolor
\endpicture
\caption{An unfolded walk from the origin in the slab $S_w$ of width $w$ 
defined by all vertices $j$ with $0 \leq x_2(j) \leq w$.  The $x_3$
direction is normal to the plane of the page.}
\label{fig2}
\end{figure}

\subsection{Confined walks}

In addition to the above results, we shall also need a result on walks 
confined to \textit{slabs}.

The situation we consider is illustrated in figure \ref{fig2}.  A 
positive walk from the origin in the half space $x_3\geq 0$ and 
adsorbing in the $x_3=0$ plane is also confined to a slab $S_w$ 
with boundaries being the two planes $x_2 = 0$ and $x_2=w$.  That is, 
this walk is adsorbing in the plane $x_3=0$ (namely the $(x_1,x_2)$-plane 
shown in the figure) and is confined in the slab with $x_3\geq 0$ 
between the two planes $x_2=0$ and $x_2=w$ normal to the $x_2$-direction 
and a distance $w$ apart (these two planes project to the horizontal 
lines bounding the slabe in figure \ref{fig2}).  Denote the number of
these walks of length $n$ by $c_n^{(w)}$. Using the methods in references 
\cite{HammersleyWhittington,Soteros88,WhittingtonSoteros1991,
WhittingtonSoteros1992} it can be shown that the limit
\begin{equation}
\lim_{n\to\infty} \frac{1}{n} \log c_n^{(w)} = \log \mu^{(w)}
\label{eq:slabgrowth}
\end{equation}
exists, and that $\mu^{(w)}<\mu^{(w+1)}<\mu_3$ and 
$\lim_{w\to\infty} \mu^{(w)} = \mu_3$.

We extend these results to walks adsorbing in the $x_3=0$ plane below.
The partition function of these walks is then
\begin{equation}
C_n^{(w)} (a) = \sum_v c_n^{(w)} (v)\, a^v ,
\end{equation}
where $c_n^{(w)}(v)$ is the number of walks with $v$ vertices in the 
plane $x_3=0$ (apart from the origin), and confined to the slab $S_w$.

Next, consider the walks in $C_n^{(w)}(a)$ which are unfolded 
in the $x_1$-direction (that is, $x_1(0) < x_1(j) \leq x_1(n)$ 
for $1 \leq j \leq n$ and $0 \leq x_2(k) \leq w$ for $0\leq k \leq n$). 
Denote the number of these unfolded walks of length $n$ in $S_w$ 
having $v$ vertices (apart from the origin) in the adsorbing plane
$x_3=0$ by $c_n^{\dagger[1],(w)}(v)$, and the corresponding partition 
function by $C_n^{\dagger[1],(w)}(a)$.

In addition, let $L_n^{\ddagger[1],(w)}(a)$ denote the partition function 
of walks contributing to $C_n^{\dagger[1],(w)}(a)$ which are also 
loops \textit{with both endpoints in the $x_1$-axis}.  That is,  
$x_1$-unfolded walks in the slab with $x_j(0)=x_j(n)=0, j=2,3$.
In this case, the double dagger superscript does not denote double 
unfolding but instead  denotes that the endpoints of the loops 
are restricted to the surface $x_3=0$ and also to the line 
$x_2=x_3=0$ (namely the $x_1$-axis). This subset of unfolded 
loops will be used in constructions later in the paper.

We have the following lemma.
\begin{lemm}
The limit
$\displaystyle \lim_{n\to\infty} \Sfrac{1}{n} \log C_n^{(w)} (a) 
= \kappa^{(w)}(a)$ exists.  Moreover, $\kappa^{(w-1)}(a) 
\leq \kappa^{(w)}(a) < \kappa(a)$, $\kappa^{(w)}(1)=\log \mu^{(w)}$ 
and $\displaystyle\lim_{w\to\infty} \kappa^{(w)}(a) = \kappa(a)$.

The same results hold if the walks in $S_w$ are restricted
to the quarter space $x_1,x_3\geq 0$ and/or restricted to loops or 
$x_1$-unfolded loops with or without the endpoints restricted 
to the $x_1$-axis.  
\label{T1}
\end{lemm}

\Pr 
Unfolded loops with both endpoints in the $x_1$-axis contributing to
the partition function $L_n^{\ddagger[1],(w)}(a)$ can be concatenated 
with unfolded loops in the partition function $L_m^{\ddagger[1],(w)}(a)$ 
by placing the first vertex of the second loop on the last vertex of the
first loop. This shows that 
$$L_m^{\ddagger[1],(w)}(a)\, L_n^{\ddagger[1],(w)}(a) 
\le L_{m+n}^{\ddagger[1],(w)}(a)$$
and this establishes the existence of the limit 
$\lim_{n\to\infty} n^{-1} \log L_n^{\ddagger[1],(w)}(a)$.

By inclusion, $L_n^{\ddagger[1],(w)}(a) \le L_n^{\dagger[1],(w)}(a)\leq 
L_n^{(w)}(a) \le C_n^{(w)} (a)$.  Similarly, $C_n^{\dagger[1],(w)}(a)$ 
and the partition functions for  walks or loops in $S_w$ and 
restricted to the quarter space $x_1,x_3\geq 0$ fit between these 
two extremes. 

Consider the subset of walks from the origin, confined to $S_w$, 
unfolded in the $x_1$-direction, with partition function 
$C_n^{\dagger[1],(w)}(a)$.  There is a subset of these walks with
most popular $(x_2,x_3)$-coordinates of their end-point. Such walks 
can be concatenated with a reverse of such a walk (and an additional 
edge in the $x_1$-direction), to form a loop with endpoints in the
$x_1$-axis.  This gives the bound
$$ \left( C_n^{\dagger[1],(w)} (a) \right)^2
\leq \max(1,a)\,(n+1)^2(w+1)^2\, L_{2n+1}^{\ddagger[1],(w)}(a) .$$
Since $L_{n}^{\ddagger[1],(w)}(a) \leq C_n^{\dagger[1],(w)}(a)$
and $C_n^{\dagger[1],(w)}(a) \le C_n^{(w)}(a) 
\le C_n^{\dagger[1],(w)}(a)\, e^{O(\sqrt{n})}$
this completes the proof that the limits 
$$\lim_{n\to\infty} \Sfrac{1}{n} \log L_n^{\ddagger[1],(w)}(a)
=\lim_{n\to\infty} \Sfrac{1}{n} \log C_n^{(w)} (a) = \kappa^{(w)}(a)$$
exist.

Setting $a=1$ turns off the interaction with the $x_3=0$ plane, 
it follows that $\kappa^{(w)}(1)=\log \mu^{(w)}$. It follows by 
inclusion that $\kappa^{(w-1)}(a) \leq \kappa^{(w)}(a)$.

Since any subwalk with $x_2$-span greater than $w$ cannot occur 
as a subwalk of any walk in the slab $S_w$,  it follows from the 
pattern theorem for adsorbing half-space walks 
(see \cite[section 7.2]{Rensburg2015}) that $\lim_{n\to\infty} 
\Sfrac{1}{n} \log C_n^{(w)}(a) = \kappa^{(w)}(a) < \kappa(a)$.

To prove that $\lim_{w\to\infty} \kappa^{(w)}(a)=\kappa(a)$, 
consider, $L_m^{\ddagger[1]}(a)$, the partition function for 
unconfined $x_1$-unfolded loops with end points in the $x_1$-axis. 
Its limiting free energy is $\kappa(a)$. Thus for any $\epsilon > 0$ 
there exists $N=N(\epsilon)$ such that $\Sfrac{1}{m} 
\log L_m^{\ddagger[1]}(a) > \kappa(a) - \epsilon$ for $m \ge N$.  
These loops will fit into a slab with $w > m$.  Fix $w > m \geq N$ 
and then put $n=mp+q$ with $n \ge N$, non-negative integers $p$ 
and $q$, and $0\leq q < m$.  By concatenating $p$ unfolded loops 
of size $m$ and a final loop of size $q$,
$$ \Sfrac{pm}{pm+q}\left( \kappa(a) - \epsilon \right)
\leq \Sfrac{1}{pm+q} \, p \log L_m^{\ddagger[1]}(a) 
+ \log L_q^{\ddagger[1]}(a)
\leq \Sfrac{1}{n} \log C_n^{(w)}(a) .$$
Take $n\to\infty$ by taking $p\to\infty$ to obtain
$\kappa(a)-\epsilon \leq \kappa^{(w)}(a)$.  This completes the proof. \qed

\section{Uniform combs}
\label{sec:combs}

\subsection{Bounds on the free energy of uniform combs}

In this section we are concerned with uniform combs.  See figure \ref{F2} 
for a sketch.  A comb can be thought of as a self-avoiding walk making up 
the backbone of the comb, with $t$ teeth attached at regular intervals along 
the backbone. The teeth are also self-avoiding walks, and the teeth and backbone
are mutually avoiding. The comb is placed in the half lattice ${\mathbb L}^3$ 
which is that part of the cubic lattice with non-negative $x_3$ coordinates.  
The first vertex of the backbone of the comb is grafted at the origin, 
and all other vertices are in ${\mathbb L}^3$.

A comb with $t$ teeth has $t$ vertices of degree 3 and $t+2$ vertices of 
degree 1. The number of edges in each subwalk between two adjacent vertices 
of degree 3 and between the end vertices of degree 3 and the initial and 
terminal vertices of degree 1 is $m_b$.  The number of edges in each tooth 
is $m_a$.  The total number of edges in the comb is $n=(t+1)\,m_b + t\,m_a$.

In the uniform case we take $m_a=m_b=m$ so that $n=(2t+1)\,m$. In this 
section we consider $t$-combs where the number of teeth $t\geq 1$ is 
a fixed integer and define 
\begin{equation}
k_n(v_A,v_B,h)=k^{(t)}(n/(2t+1),n/(2t+1),v_A,v_B,h)
\end{equation}
to be the number of uniform $t$-combs with a total of $n$ edges, 
having $v_A$ $A$-visits and $v_B$ $B$-visits, 
and with the 
$x_3$-coordinate of the terminal vertex of the backbone equal to $h$.
The corresponding partition function in terms of the general 
partition function   (\ref{eq:genpf}) is 
\begin{equation}
\fl K_n(a,b,y) = K^{(t)}(n/(2t+1),n/(2t+1),a,b,y)
=\sum_{v_A,v_B,h}k_n(v_A,v_B,h)\, a^{v_A}b^{v_B} y^h.
\label{eqn:uniformPF}
\end{equation}

\begin{figure}[h!]
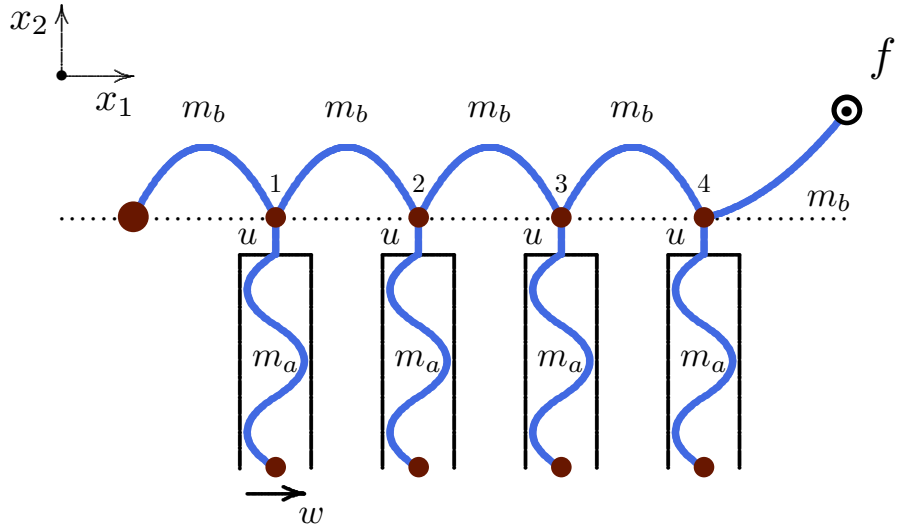

\beginpicture
\setcoordinatesystem units <1.35pt,1.35pt> 
\setplotarea x from -80 to 200, y from -70 to 60

\color{Black}
\normalcolor

\arrow <7pt>  [.2,.67] from -20 40 to 0 40 
  \put {$\scalebox{1.5}{$x_1$}$} at -5 32
\arrow <7pt>  [.2,.67] from -20 40 to -20 60 
  \put {$\scalebox{1.5}{$x_2$}$} at -29 55
\put {$\bullet$} at -20 40 

\setplotsymbol ({\scalebox{0.33}{$\bullet$}})
\color{Black}
\setdots <4pt>
\plot -20 0 200 0 /
\setsolid
\plot 30 -70 30 -10 50 -10 50 -70 /
\plot 70 -70 70 -10 90 -10 90 -70 /
\plot 110 -70 110 -10 130 -10 130 -70 /
\plot 150 -70 150 -10 170 -10 170 -70 /

\setdots <2pt>
\setplotsymbol ({\scalebox{0.33}{$\bullet$}})
\setsolid
\arrow <8pt> [.2,.67] from 32 -77 to 48 -77 

\setplotsymbol ({\scalebox{0.67}{$\bullet$}})
\color{RoyalBlue}
\setquadratic 
\plot 0 0 20 20 40 0  /
\plot 40 0 60 20 80 0 /
\plot 80 0 100 20 120 0 /
\plot 120 0 140 20 160 0 /
\plot 160 0 180 10 200 30 /
\setlinear
\plot 40 0 40 -10 /
\plot 80 0 80 -10 /
\plot 120 0 120 -10 /
\plot 160 0 160 -10 /

\setquadratic
\plot 40 -10 32 -20 40 -30 48 -40 40 -50 32 -60 40 -70 /
\plot 80 -10 72 -20 80 -30 88 -40 80 -50 72 -60 80 -70 /
\plot 120 -10 112 -20 120 -30 128 -40 120 -50 112 -60 120 -70 /
\plot 160 -10 152 -20 160 -30 168 -40 160 -50 152 -60 160 -70 /

\color{Sepia}
\multiput {\scalebox{2.00}{$\bullet$}} at 40 0 80 0 120 0 160 0
40 -70 80 -70 120 -70 160 -70 /
\multiput {\scalebox{3.00}{$\bullet$}} at 0 0 /

\color{black}
\put {\scalebox{3}{$\bullet$}} at 200 30
\color{white}
\put {\scalebox{2}{$\bullet$}} at 200 30 
\color{black}
\put {\scalebox{1}{$\bullet$}} at 200 30

\put {\scalebox{1.75}{$f$}} at 210 45

\multiput {\scalebox{1.33}{$m_b$}} at 20 30 60 30 100 30 140 30 195 5 /
\multiput {\scalebox{1.33}{$u$}} at 32 -5 72 -5 112 -5 152 -5 /
\multiput {\scalebox{1.33}{$m_a$}} at 40 -40 80 -40 120 -40 160 -40 /
\multiput {\scalebox{1.33}{$w$}} at 50 -83 /
\put {1} at 40 10 \put {2} at 80 10 \put {3} at 120 10 \put {4} at 160 10
\normalcolor
\endpicture
\caption{Top view schematic projected onto the $x_1x_2$-plane of an 
adsorbed comb in the half-space $x_3\geq 0$ grafted at the origin (on the left)
and pulled by a vertical force at its endpoint (on the right) from the adsorbing
plane.  The backbone consists of unfolded loops with endpoints in the $x_1$-axis
(dotted line) in the half-space $x_2\geq 0$.  The teeth are confined in slabs
of width $w$ normal to the $x_1$-axis in the half-space $x_2<0$.}
\label{fig4}
\end{figure}

\begin{figure}[h!]
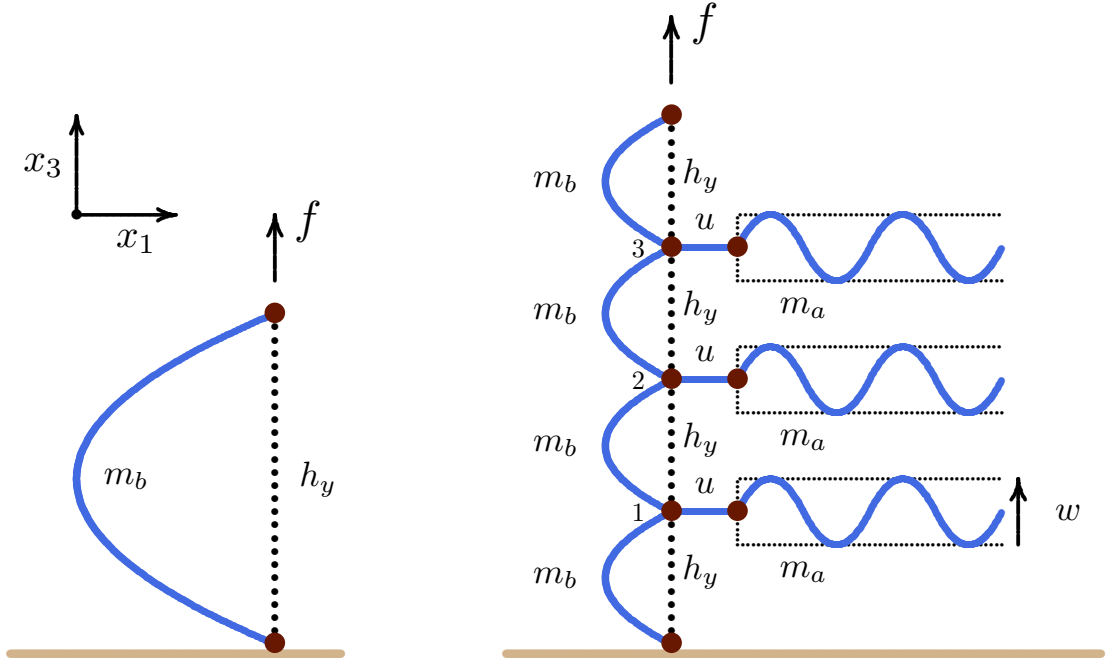

\beginpicture
\setcoordinatesystem units <1.25pt,1.25pt> 
\setplotarea x from -10 to 100, y from -10 to 200

\setplotsymbol ({\scalebox{0.8}{$\bullet$}})
\color{Tan}
\plot 0 -3  100 -3 /

\setplotsymbol ({\scalebox{0.33}{$\bullet$}})
\color{Black}
\arrow <8pt> [.2,.67] from 80 110 to 80 130  
  \put {\scalebox{1.75}{$f$}} at 90 128
\arrow <7pt>  [.2,.67] from 20 130 to 50 130 
  \put {$\scalebox{1.5}{$x_1$}$} at 38 122
\arrow <7pt>  [.2,.67] from 20 130 to 20 160 
  \put {$\scalebox{1.5}{$x_3$}$} at 10 145
\put {$\bullet$} at 20 130

\setplotsymbol ({\scalebox{0.67}{$\bullet$}})
\color{Black}
\setdots <5pt>
\plot 80 0 80 100 /

\setsolid
\setplotsymbol ({\scalebox{0.67}{$\bullet$}})
\color{RoyalBlue}
\setquadratic 
\plot 80 0 20 50 80 100 /

\setlinear
\color{Black}
\put {\scalebox{1.33}{$m_b$}} at 35 50 
\put {\scalebox{1.33}{$h_y$}} at 93 50 

\color{Sepia}
\multiput {\scalebox{2}{$\bullet$}} at 80 0 80 100 /

\setcoordinatesystem units <1.25pt,1.25pt> point at -150 0 
\setplotarea x from -30 to 200, y from -10 to 200

\setplotsymbol ({\scalebox{0.8}{$\bullet$}})
\color{Tan}
\plot 0 -3  180 -3 /

\setplotsymbol ({\scalebox{0.67}{$\bullet$}})
\color{Black}
\setdots <5pt>
\plot 50 0 50 160 /

\setdots <2pt>
\setplotsymbol ({\scalebox{0.33}{$\bullet$}})
\plot 150 30 70 30 70 50 150 50 /
\plot 150 70 70 70 70 90 150 90 /
\plot 150 110 70 110 70 130 150 130 /
\setsolid
\arrow <8pt> [.2,.67] from 155 30 to 155 50

\setplotsymbol ({\scalebox{0.67}{$\bullet$}})
\color{RoyalBlue}
\setquadratic 
\plot 50 0 30 20 50 40 /
\plot 50 40 30 60 50 80 /
\plot 50 80 30 100 50 120 /
\plot 50 120 30 140 50 160 /
\setlinear
\plot 50 40 70 40 / \plot 50 80 70 80 / \plot 50 120 70 120 /

\setquadratic
\plot 70 40 80 50 90 40 100 30 110 40 120 50 130 40 140 30 150 40 /
\plot 70 80 80 90 90 80 100 70 110 80 120 90 130 80 140 70 150 80 /
\plot 70 120 80 130 90 120 100 110 110 120 120 130 130 120 140 110 150 120 /

\color{Sepia}
\multiput {\scalebox{2.00}{$\bullet$}} at 50 0 50 40 50 80 50 120 50 160 
70 40 70 80 70 120 /

\color{black}
\setplotsymbol ({\scalebox{0.33}{$\bullet$}})
\arrow <8pt> [.2,.67] from 50 170 to 50 190  
  \put {\scalebox{1.75}{$f$}} at 60 188

\multiput {\scalebox{1.33}{$m_b$}} at 15 20 15 60 15 100 15 140 /
\multiput {\scalebox{1.33}{$u$}} at 60 48 60 88 60 128 /
\multiput {\scalebox{1.33}{$m_a$}} at 90 22 90 62 90 102 /
\multiput {\scalebox{1.33}{$h_y$}} at 59 22 59 62 59 102 59 142 /
\multiput {\scalebox{1.33}{$w$}} at 170 40 /
\put {1} at 40 40 \put {2} at 40 80 \put {3} at 40 120 
\normalcolor

\endpicture
\caption{(Left) A schematic drawing of a pulled unfolded loop.  The 
loop is oriented in the half-space $x_1\leq 0$ and its endpoints are 
in the plane $x_1=0$.  A force $f$ is applied in the vertical direction.  
If the length of the loop is $m_b$, then there is a most popular 
vertical separation (or height) $h_y$ between its endpoints.  This 
most popular height is a function of both $m_b$ and $y$.
(Right) Constructing a pulled comb by combining pulled loops with 
teeth confined to horizontal slits or slabs of (constant) height $w$.  
In this case, each loop has length $m_b$ and height $h_y$ (the most 
popular height at an applied force $f$).  Each tooth has length $m_a$ 
and is joined to the backbone by a $u$ horizontal edges.}
\label{figg3}
\end{figure}

In order to prove existence of a unique limiting free energy 
$\zeta(a,b,y) = \lim_{n\to\infty} \Sfrac{1}{n} \log K_n(a,b,y)$
in this model, we shall consider the schematic conformations of adsorbed
combs shown in figures \Ref{fig4} and \Ref{figg3}.  The proof proceeds
by establishing lower and upper bounds on $K_n(a,b,y)$.

We proceed by first considering $x_1$-unfolded adsorbing $B$-loops 
and denote their partition function by $L_n^{\dagger[1]}(b)$.  In 
figure \ref{fig4} we are interested in such loops along the backbone 
of the comb with $x_1$-spans greater than a constant value $w$, and 
we denote the partition function of these by $L_n^{\dagger[1]}({>}w,b)$.  
The remaining loops of $x_1$-span less than or equal to $w$ have 
partition function $L_n^{\dagger[1]}({\leq}w,b)$.  Then it follows that
\begin{equation}
L_n^{\dagger[1]}(b) = L_n^{\dagger[1]}({>}w,b) 
+ L_n^{\dagger[1]}({\leq}w,b).
\label{10a}
\end{equation}
A corresponding equation arises for the partition functions for 
subsets of these loops that are restricted further either to 
$x_2 \geq 0$ and/or to having end points in the $x_1$-axis.  
We need the following lemma in order to find a lower bound on
the free energy of uniform combs.

\begin{lemm}
For any fixed $w\geq 0$, $\displaystyle 
\lim_{n\to\infty} \Sfrac{1}{n} \log L_n^{\dagger[1]}({>}w,b)
= \lim_{n\to\infty} \Sfrac{1}{n} \log L_n^{\dagger[1]}(b)
= \kappa(b)$.
The same result holds for the subsets of these loops with $x_2\geq 0$ 
and/or with the endpoints restricted to the $x_1$-axis.  
\label{lemma3}
\end{lemm}
\Pr Take logarithms on the left hand side of equation (\Ref{10a}), 
divide by $n$, and let $n\to\infty$.  This gives
for this case (and for the other restrictions considered) 
\[ \kappa(b) 
= \lim_{n\to\infty} \Sfrac{1}{n}
\log L_n^{\dagger[1]}(b)
= \lim_{n\to\infty} \Sfrac{1}{n}
\log \left( L_n^{\dagger[1]}({>}w,b) 
 + L_n^{\dagger[1]} ({\leq}w,b) \right) .\]
The walks contributing to $L_n^{\dagger[1]} ({\leq}w,b)$ correspond 
to subsets of walks that fit in the slab of width $w$, $S_w$, as 
in figure \ref{fig2} except now the restricting planes are 
constant $x_1$ planes instead of constant $x_2$ planes. By symmetry, 
the results of lemma \ref{T1} also hold for walks in $S_w$ 
(or with the additional restriction $x_2\geq 0$).  It is thus a 
corollary of lemma \ref{T1} that
\[ \limsup_{n\to\infty} \Sfrac{1}{n} \log L_n^{\dagger[1]} ({\leq}w,b)
\leq \max_{v\leq w} \kappa^{(v)}(b) < \kappa(b).\]
Existence of the limits above then implies that
$\lim_{n\to\infty} \Sfrac{1}{n}\log  L_n^{\dagger[1]}({>}w,b)
=\kappa(b)$. \qed  

Next, consider figure \ref{figg3}. On the left hand side is a 
negative $x_1$-loop (that is, a loop oriented in the $x_1$ direction in 
the half-space $x_1\leq 0$). It is $x_3$-unfolded and pulled by 
a force in the $x_3$-direction.  These loops have the same partition 
function and free energy, $\lambda(y)$ for $y>0$, as the unfolded 
loops of Lemma \ref{lemm:lemma1}. Suppose that the length of the 
loop is $m_b$.  Then there is a most popular (vertical) distance 
or \textit{height} between its endpoints, denoted by $h_y$ as in 
the figure (this is a function of the pulling force $f= k_BT \log y$ 
and of the length $m_b$). The number of unfolded loops of length 
$m_b$ and height $h_y$ is denoted by $\ell_{m_b}^{\dagger[3]}(h)$, 
and it follows that
\begin{equation}
[ \ell_{m_b}^{\dagger[3]}(h_y)\,y^{h_y} ]
\leq \sum_{h=0}^{m_b} \ell_{m_b}^{\dagger[3]} (h)\, y^h
\leq (m_b+1) [ \ell_{m_b}^{\dagger[3]}(h_y)\,y^{h_y} ].
\label{eq:popheight}
\end{equation}
Since, by Lemma \ref{lemm:lemma1}, the free energy of these pulled 
unfolded loops exists and is equal to $\lambda(y)$, this shows that
\begin{equation}
\lim_{m_b\to\infty} \Sfrac{1}{m_b} 
\log [ \ell_{m_b}^{\dagger[3]} (h_y)\,y^{h_y} ] = \lambda (y) .
\label{eq:limhy}
\end{equation}
It is the case that $h_y$ is not bounded, but grows with $m_b$.  
This is shown in the next lemma.

\begin{lemm}
$\displaystyle\liminf_{m_b\to\infty} h_y = \infty$ for all $y>0$.
\label{lemma2a}
\end{lemm}
\Pr
Suppose that there exists a $w\geq 0$ such that $h_y < w$ for 
all $m_b\geq 0$.  Then $\ell_{m_b}^{\dagger[3]}(h_y) 
\leq c_{m_b}^{(w)}$ (that is, each loop of length $m_b$ and height 
$h_y$ is also a self-avoiding walk confined in a slab of width $w$, 
bounded by the planes $x_3=0$ and $x_3=w$).  Taking logarithms, 
dividing by $m_b$ and then taking $m_b\to\infty$ gives via 
equation (\ref{eq:slabgrowth}), for $y > 0$,
\[ \fl
\lambda(y) = \lim_{m_b\to\infty} \Sfrac{1}{m_b}
 \log [\ell_{m_b}^{\dagger[3]} (h_y)\,y^{h_y} ]
\leq \lim_{m_b\to\infty} \Sfrac{1}{m_b}\log [ c_{m_b}^{(w)} \,
(\max\{1,y\})^w] = \log \mu^{(w)} < \log \mu_3 .\]
This is a contradiction (since $\lambda(y)\geq \log \mu_3$), and 
thus $h_y$ is not bounded.  Since the limit on the left exists, 
this is valid if the limits are taken along any subsequence, 
with the result that $h_y$ is unbounded along any subsequence.  
Thus, for any $N>0$ there is an $M$ such that $h_y>N$ for all 
$m_b>M$. \qed

\begin{theo}
When $y \ge 1$ the free energy for uniform $t$-combs
$\displaystyle
\zeta(a,b,y)=  \lim_{n\to\infty} \Sfrac{1}{n} \log K_n(a,b,y)$ 
is given by
$$\zeta(a,b,y)=\max \left[\Sfrac{t}{2t+1}\kappa(a) 
+ \Sfrac{t+1}{2t+1} \kappa(b), \Sfrac{1}{2t+1}\lambda(y) 
+ \Sfrac{t}{2t+1}(\kappa(a)+\kappa(b)),
\Sfrac{t+1}{2t+1} \lambda(y) + \Sfrac{t}{2t+1} \log \mu_3
\right].$$
\label{theo:uniformFE}
\end{theo}
\Pr \\
\noindent\underbar{Lower bound:}
First establish a lower bound by considering $t$-combs in conformations
shown in figures \ref{fig4} and \ref{figg3}.  We label the trivalent
nodes in the combs, starting in the node closest to the origin by
$1$, $2$,\ldots, $t$, as shown. A lower bound is constructed by 
considering three cases, based on the schematic diagrams:
\begin{enumerate}
\item[1)] the height of the pulled vertex is 0 (a special case of figure \ref{fig4});
\item[2)] the height of trivalent vertex labeled $t$ is 0 and  all the vertices 
in the $(t+1)$-st branch are out of the surface (see figure \ref{fig4});
\item[3)] all vertices of the comb are above the surface except for the origin 
(see figure \ref{figg3}).
\end{enumerate}

Consider case (1) and figure \ref{fig4} but with the last (pulled)
vertex at height zero.  The backbone of the comb is a
sequence of $t$ loops unfolded in the $x_1$-direction, each of length
$m_b$ and confined to the sublattice $x_2\geq 0$ and $x_3\geq 0$, with
endpoints in the $x_1$-axis.  The last segment along the backbone is
also assumed to be an unfolded loop with endpoints in the $x_1$-axis.

Next, assume that the $x_1$-span of each loop along the backbone 
exceeds $w$.  We simplify the presentation by denoting the partition
function for each of these loops by $L_n^{\dagger[1]}({>}w,b)$ 
(this includes the  restrictions that $x_2\geq 0$ and the loop 
endpoints are in the $x_1$-axis). Thus by lemma \ref{lemma3} their 
limiting free energies are $\kappa(b)$.

The teeth in figure \ref{fig4} are quarantined in slabs of width
$w$ (bounded by constant $x_1$ planes) in the sublattice $x_2<0$ 
and $x_3\geq 0$, each of length $m_a$, and interacting with the 
$x_3$-plane.  Each tooth is attached to the backbone by a sequence 
of $u$ edges in the $x_2$ direction, and one may assume that $u=1$.  
By symmetry, these walks have the same partition function as the 
quarter space walks of  lemma \ref{T1}. For simplicity, we denote 
that partition function  here  by  $C_{m_a-u}^{(w)}(a)$ (this
includes the quarter space restriction). By lemma \ref{T1} its 
limiting free energy is $\kappa^{(w)}(a)$. 

Thus, the comb is now put together by assuming each tooth is
quarantined in a slab of width fixed at $w$, and the loops along
the backbones have $x_1$-spans exceeding $w$. This shows that
for case (1) above, 
\begin{eqnarray}
    K^{(t)}(m_a,m_b,a,b,y) \geq 
\left[L_{m_b}^{\dagger[1]}({>}w,b)\right]^{t} \;
\left[L_{m_b}^{\dagger[1]}(b)\right]^{1} \;
\left[ C_{m_a-u}^{(w)}(a)\right]^t .  
\label{eq:15}
\end{eqnarray}
Next, put $m_a=m_b=m$, take logarithms, divide by $(2t+1)\,m$, 
and then take $m\to\infty$.  By lemmas \ref{T1} and \ref{lemma3} 
and equation (\ref{eq:psidef}),
\begin{equation}
 \liminf_{m\to\infty} \Sfrac{1}{(2t+1)\,m} \log
 K^{(t)}(m,m,a,b,y) \geq \Sfrac{1}{(2t+1)}(t\kappa(b) 
 + \kappa(b) + t\kappa^{(w)}(a) ).
\end{equation}
Since $w$ is arbitrary, one may take $w\to\infty$ on the 
right hand side.  By lemma \ref{T1} $\kappa^{(w)}(a)\to \kappa(a)$.
This gives the lower bound $\Sfrac{t}{2t+1}\kappa(a) 
+ \Sfrac{t+1}{2t+1} \kappa(b)$ from case (1).

Case (2) is treated in the same way, except that the last segment
of the backbone is anchored by the $t$-th trivalent vertex to the 
adsorbing plane $x_3=0$ but is otherwise disjoint from it, and is
pulled in its endpoint.  This gives the lower bound
\begin{equation}
 \liminf_{m\to\infty} \Sfrac{1}{(2t+1)\,m} \log
 K^{(t)}(m,m,a,b,y) \geq \Sfrac{1}{2t+1}(t\kappa(b) 
 + \lambda(y) + t\kappa^{(w)}(a)).
 \label{eq:17}
\end{equation}
Taking $w\to\infty$ gives the lower bound 
$\Sfrac{1}{2t+1}\lambda(y) + \Sfrac{t}{2t+1}(\kappa(a)+\kappa(b))$ 
from case (2).

Consider case (3) next.  The proof is again similar to that of case (1),
but now loops are stacked in the $x_3$-direction as shown in figure
\ref{figg3}. Let $y>0$, fix $m_b$ and choose $w<h_y$, with $h_y$ as 
defined before equation (\ref{eq:popheight}).  Put $u=1$.  Then
\begin{eqnarray}
    K^{(t)}(m_a,m_b,a,b,y) \geq 
\left[\ell_{m_b}^{\dagger[3]}(h_y)\,y^{h_y}\right]^{t+1} \;
\left[ c_{m_a-u}^{\dagger[1],(w)}\right]^t .
\label{eq:18}
\end{eqnarray}
Next, put $m_a=m_b=m$, take logarithms, divide by $(2t+1)\,m$, and then
take $m\to\infty$. By equation (\ref{eq:limhy}) and lemma \ref{T1},
\begin{equation}
 \liminf_{m\to\infty} \Sfrac{1}{(2t+1)\,m} \log
 K^{(t)}(m,m,a,b,y) \geq \Sfrac{1}{2t+1}
 \left( (t+1)\lambda(y) + t\,\log \mu^{(w)} \right) .
\end{equation}
By lemma \ref{lemma2a}, $h_y$ increases to infinity as $m\to\infty$,
thus $w<h_y$ can be made arbitrarily large.  Since 
$\mu^{(w)} \to \mu_3$, this completes case (3).

This completes the lower bound.

\begin{figure}[h!]
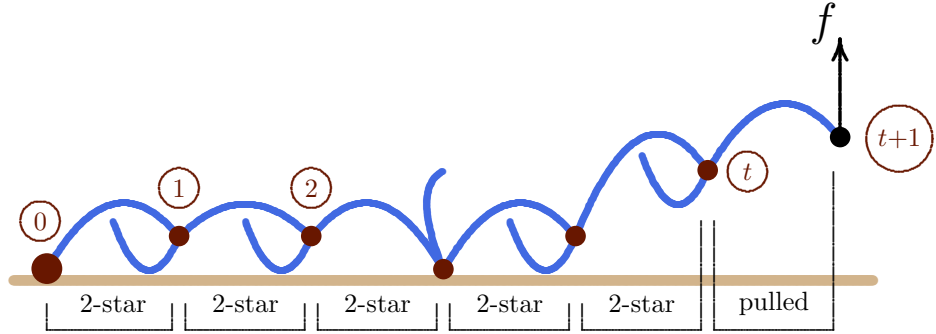

\beginpicture
\setcoordinatesystem units <1.25pt,1.25pt> 
\setplotarea x from -80 to 200, y from -25 to 80

\setplotsymbol ({\scalebox{1.00}{$\bullet$}})
\color{Tan}
\plot -20 -3 240 -3 /

\setplotsymbol ({\scalebox{0.16}{$\bullet$}})
\color{Black}
\setsolid
\plot -10 -10 -10 -18 28 -18 28 -10 /
\plot 32 -10 32 -18 68 -18 68 -10 /
\plot 72 -10 72 -18 108 -18 108 -10 /
\plot 112 -10 112 -18 148 -18 148 -10 /
\plot 152 -10 152 -18 188 -18 188 15 /
\plot 192 15 192 -18 228 -18 228 30 /

\setplotsymbol ({\scalebox{0.67}{$\bullet$}})
\color{RoyalBlue}
\setquadratic 
\plot -10 0 10 20 30 10 /
\plot 30 10 50 20 70 10 /
\plot 70 10 90 20 110 0 /
\plot 110 0 130 20 150 10 /
\plot 150 10 170 40 190 30 /
\plot 190 30 210 50 230 40 /

\setquadratic
\plot 30 10 20 0 10 15 /
\plot 70 10 60 0 50 15 /
\plot 110 0 105 20 110 30 /
\plot 150 10 140 0 130 15 /
\plot 190 30 180 20 170 35 /

\color{Sepia}
\multiput {\scalebox{2.00}{$\bullet$}} at 30 10 70 10 110 0 
150 10 190 30  /
\multiput {\scalebox{3.00}{$\bullet$}} at -10 0 /
\put {$0$} at -12 15  \put {$1$} at 30 25  \put {$2$} at 70 25
\put {$t$} at 202 30 \put{$t{+}1$} at 248 40
\setplotsymbol ({\scalebox{0.2}{$\bullet$}})
\multiput {\beginpicture
           \circulararc 360 degrees from 6 0 center at 0 0 
           \endpicture} at -12 15 30 25  70 25  202 30 /
\multiput {\beginpicture
           \circulararc 360 degrees from 10 0 center at 0 0 
           \endpicture} at 248 40 /

\setplotsymbol ({\scalebox{0.3}{$\bullet$}})
\color{black}
\arrow <8pt> [.2,.67] from 230 40 to 230 70
\put {\scalebox{2}{$\bullet$}} at 230 40 

\put {\scalebox{1.75}{$f$}} at 225 75

\multiput {$2$-star} at 10 -10 50 -10 90 -10 130 -10 170 -10 /
\multiput {pulled} at 210 -10 /
\normalcolor
\endpicture
\caption{An adsorbing and pulled uniform comb. A force $f$ is pulling 
at the end of the backbone, desorbing the comb from the surface.
Cutting the comb in its trivalent nodes gives a sequence of copolymeric 
$2$-stars and a final branch where the comb is pulled. The trivalent
nodes of the comb are labelled starting at the origin by $0$, $1$,
$\ldots$, $t$, $t+1$.}
\label{fig55}
\end{figure}

\vspace{2mm}
\noindent\underbar{Upper bound:}

\noindent{{\bf The case $\mathbf{y\geq 1}$}:} 
The bound is constructed by viewing the comb as composed of a sequence
of copolymer $2$-stars, followed by a final pulled branch on the backbone
as shown schematically in figure \ref{fig55}.  The pulling force $f$
is transmitted along the backbone to the origin, or along the backbone
via an absorbed trivalent vertex or a tooth, or an adsorbed branch along 
the backbone, into the adsorbing plane.

We construct an upper bound by considering the $2$-stars to be 
independent of each other, and bound each from above.

As before, assume that $K^{(t)}(m_a,m_b,a,b,y)$ is the partition function
of the comb, where the backbone branches have lengths $m_b$, the teeth
have lengths $m_a$, and $(a,b,y)$ are the parameters as defined before.

\begin{figure}[h]
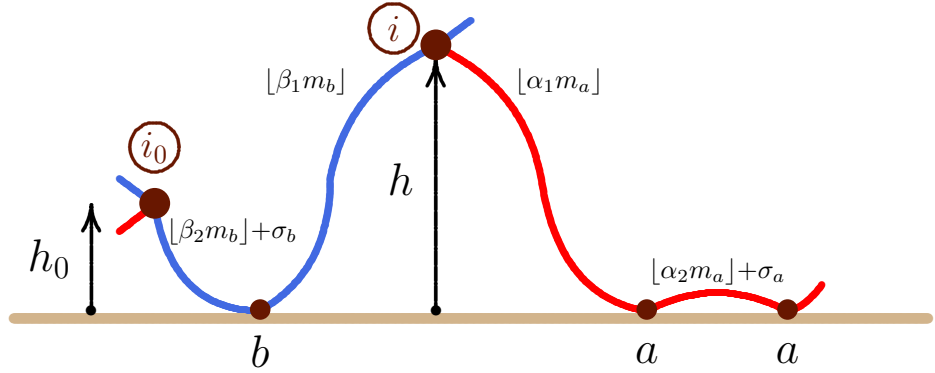

\beginpicture
\setcoordinatesystem units <1.33pt,1pt>
\setplotarea x from -75 to 200, y from -10 to 100

\setplotsymbol ({\scalebox{1.00}{$\bullet$}})
\color{Tan}
\plot -20 -3 240 -3 /

\setplotsymbol ({\scalebox{0.32}{$\bullet$}})
\color{Black}
\setsolid

\arrow <10pt> [.2,.67] from 2 0 to 2 40 
\put {$\bullet$} at 2 0 
\put {\LARGE$h_0$} at -10 20 

\arrow <10pt> [.2,.67] from 100 0 to 100 95 
\put {$\bullet$} at 100 0 
\put {\LARGE$h$} at 90 50 

\put {\LARGE$b$} at 50 -15
\multiput {\LARGE$a$} at 160 -16 200 -16 /

\color{Sepia}
\put {\Large$i$} at 88 107
\put {\Large$i_0$} at 20 62
\setplotsymbol ({\scalebox{0.3}{$\bullet$}})
\multiput {\beginpicture
           \circulararc 360 degrees from 7 0 center at 0 0 
           \endpicture} at 88 107  20 62 /

\setplotsymbol ({\scalebox{0.67}{$\bullet$}})
\color{RoyalBlue}
\setquadratic 
\plot 100 100 80 80 70 50 65 20 50 0 30 10 20 40  /
\setlinear
\plot 20 40 10 50 /
\plot 100 100 110 110 /

\color{Red}
\setquadratic
\plot 100 100 120 80 130 50 140 15 160 0 170 5 180 7 190 5 200 0 205 3 210 10 /
\setlinear
\plot 20 40 10 30 /

\color{Sepia}
\multiput {\scalebox{2.00}{$\bullet$}} at 50 0 160 0 200 0 /
\multiput {\scalebox{3.00}{$\bullet$}} at 20 40 100 100 /

\color{black}
\put {$\lfloor\beta_2m_b\rfloor{+}\sigma_b$} at 42 30
\put {$\lfloor\beta_1m_b\rfloor$} at 63 87
\put {$\lfloor\alpha_1m_a\rfloor$} at 135 87
\put {$\lfloor\alpha_2m_a\rfloor{+}\sigma_a$} at 180 15

\normalcolor
\endpicture
\caption{A block copolymeric $2$-star in the comb.  The trivalent
node labelled $i$ is the central node of the star, and $i_0=i-1$ is the
central node of the preceding $2$-star. The arm of the $2$-star from $i_0$ 
to $i$ has length $m_b$ and is in the backbone of the comb.  The pulling 
force on the comb is transmitted along this arm from $i$ to $i_0$.  
The $i$-th tooth of the comb is shown on the right, and it is the other 
arm of the $2$-star.  It has length $m_a$ and adsorbs in the adsorbing plane 
with activity $a$.  Vertices in the backbone adsorb with activity $b$.  The 
lengths of parts of the arms are shown.  Here, $\sigma_a,\sigma_b\in\{0,1\}$.}
\label{fig56}
\end{figure}

Next, consider one of the $2$-stars along the comb (as shown in figure 
\ref{fig55}).  A schematic diagram of such a $2$-star is shown in 
figure \ref{fig56}.  The $2$-star has arms of lengths equal to $m_a$ and 
$m_b$.  Each arm is partitioned into a part from the central trivalent node 
(with label denoted by $i$ in figure \ref{fig56} and preceded by the trivalent 
node $i_0=i-1$) to its \textit{first visit} in the adsorbing surface, and then 
a remaining part.  The remaining part intersects the adsorbing surface, and 
those visits are weighted as shown.  

The trivalent node $i_0=i-1$ in figure \ref{fig56} is the central node 
of the preceding $2$-star, and it has height $h_0$ above the adsorbing plane.
$h_0$ is bounded by $0\leq h_0\leq(t+1)\,m_b=H$.  The trivalent node 
$i$ in figure \ref{fig56} has height $h$ above the adsorbing surface.  
Clearly, $0\leq h\leq H$.  The pulling force on the comb is transmitted to
$i$ and then passed to the adsorbing plane at the visits $a$ or $b$,  or 
to the node $i_0$.  Denote the height of the $i$-th trivalent
node by $h^{(i)}$ and the height of the first vertex of the $i$-the star by
$h_0^{(i)}\equiv h^{(i-1)}$.  The height of the entire comb is $h^{(t+1)}$
and the height of the final pulled walk is $h^{(t+1)}-h^{(t)}$.  Then
\[ \sum_{i=1}^{t+1} (h^{(i)}-h^{(i-1)}) = h^{(t+1)}-h^{(0)} = h^{(t+1)} \]
which is the height of the comb. In the partition function of the $i$-th $2$-star
the height is defined as $h^{(i)}-h_0^{(i)}=h^{(i)}-h^{(i-1)}$ and this is
conjugate to $y$.

The arm of the $2$-star which is also a branch along the backbone of the
comb (the blue arm in figure \ref{fig56}) has an \textit{adsorbing part} of 
length $\lfloor \beta_2 m_b\rfloor+\sigma_b$ and a \textit{pulled part} 
of length $\lfloor \beta_1 m_b\rfloor$.  Here, $\beta_1+\beta_2=1$ and 
$\sigma_b$ is a function of $(\beta_1,m_b)$ such that 
$\lfloor \beta_1 m_b\rfloor+\lfloor \beta_2 m_b\rfloor +\sigma_b=m_b$.   
Clearly, $0 \leq \sigma_b\leq 1$.

Similarly, the tooth (the red arm in figure \ref{fig56}) in the $2$-star has 
a pulled part of length $\lfloor\alpha_1m_a\rfloor$ and an adsorbing part 
of length $\lfloor\alpha_2m_a\rfloor+\sigma_a$, where $\alpha_1+\alpha_2=1$
and $\sigma_a$ is a function of $(\alpha_1,m_a)$ such that 
$\lfloor\alpha_1m_a\rfloor+\lfloor\alpha_2m_a\rfloor+\sigma_a=m_a$.
Again, $0 \leq \sigma_a\leq 1$.

Notice that if $\beta_2>0$ then $\lfloor \beta_2m_b\rfloor+\sigma_b \geq h_0$.
Similarly, $\lfloor \beta_1 m_b \rfloor \geq h$ and 
$\lfloor \alpha_1m_a\rfloor \geq h$.

If $\beta_1< 1/m_b$, then $h=0$, and thus $\alpha_1 < 1/m_a$.
Similarly, if $\alpha_1< 1/m_a$, then $h=0$ and thus $\beta_1<1/m_b$.
In particular, there is a $0 < \delta < \min (1/m_a,1/m_b)$ such that
$\beta_1<\delta \Leftrightarrow \alpha_1<\delta$.

This shows that if $m_a\to\infty$, or $m_b\to\infty$, then $\beta_1=0
\Leftrightarrow \alpha_1=0$.  

A similar approach to the above with
$\alpha_1$ and $\beta_1$ taking discrete values such that $\alpha_1m_a$
and $\beta_1m_b$ are integers can be followed instead.

Proceed by excising the $2$-star in figure \ref{fig56} and considering it
in isolation, independent from the rest of the comb.  These  $2$-stars
have a fixed end point in one arm at height $h_0$ above the adsorbing
surface, and are pulled in the midpoint by a vertical force, while the 
arms adsorb with activities $a$ and $b$.  

Copolymer $2$-stars terminally attached to the surface and pulled from 
their midpoint have been studied previously in reference \cite{Rensburg2022}.  
The $2$-star in figure \ref{fig56} is not terminally attached but rather 
the height  of the initial vertex is fixed at $h_0$ where $h_0$ can range 
from 0 to $(i-1)\,m_b\leq (t+1)\,m_b=H$ in the comb.  

Denote by $S_{m_a,m_b}^{\alpha_2\alpha_1\beta_1\beta_2}(h_0;a,b,y)$ 
the partition function of $2$-stars as in figure \ref{fig56} with 
$0\leq h_0 \leq H$ and where $y$ is conjugate to $h-h_0$.  If the 
central vertex of a 2-star is under tension due to the pulling force, 
then its contribution to the comb partition function is bounded above by 
\[ \sum_{\alpha_1m_a=0}^{m_a}
\sum_{\beta_1m_b=0}^{m_b}\sum_{h_0=0}^H 
S_{m_a,m_b}^{\alpha_2\alpha_1\beta_1\beta_2}(h_0;a,b,y). \]
The summations involving $\alpha_1$ and $\beta_1$ are over all 
values of $\alpha_1$ and $\beta_1$ such that $\alpha_1m_a$ and 
$\beta_1 m_b$ are integers in the sets $\{0,1,2,\ldots,m_a\}$ 
and $\{0,1,2,\ldots,m_b\}$  and these choices  fix the values 
of $\{\alpha_2,\beta_2,\sigma_a, \sigma_b\}$.

Denote the partition function of a pulled and adsorbing self-avoiding 
walk of length $m_b$ starting in a vertex at height $h_0$ above 
the adsorbing surface by $C_{m_b}(h_0;b,y)$. This corresponds to 
the $2$-star partition function with $m_a=0$.  The partition 
function of the comb is bounded above by the product of the 
partition functions of the contributing $2$-stars and the partition 
function of the final pulled walk.  In the upper bound these 
contributing factors are treated as independent.  Since $y\geq 1$, 
it follows that
\begin{eqnarray}
& &\hspace{-2cm} K^{(t)}(m_a,m_b,a,b,y)
\leq \left[ \sum_{\alpha_1m_a=0}^{m_a} 
\sum_{\beta_1m_b=0}^{m_b} \left[ \sum_{h_0=0}^H 
S_{m_a,m_b}^{\alpha_2\alpha_1\beta_1\beta_2}(h_0;a,b,y) \right]\right]^t 
\sum_{h_0=0}^H C_{m_b}(h_0;b,y)
\label{eqn56} \\
& & \hspace{-2cm} \leq [(m_a+1)(m_b+1)]^t(H+1)^{t+1}
\left[ \max_{\alpha_1,\beta_1,h_0} 
   S_{m_a,m_b}^{\alpha_2\alpha_1\beta_1\beta_2}(h_0;a,b,y) \right]^t 
\max_{h_0} C_{m_b}(h_0;b,y)  .
\label{eqn57}
\end{eqnarray}
The maximum over $\alpha_1$ and $\beta_1$ is over all values of 
$\alpha_1$ and $\beta_1$ between 0 and 1 and the maxima  fix the
values of $\{\alpha_2,\beta_2,\sigma_a, \sigma_b\}$.

We continue by putting $m_a=m_b$ and by examining the factors 
in equation (\ref{eqn57}).

\vspace{3mm}
\noindent{\textbf{Claim:}}  Given $\epsilon>0$, there exists an 
$M\geq 0$ such that for all $m\geq M$,
\[ 
\hspace{-2cm}
\frac{1}{m} \log S_{m,m}^{\alpha_2\alpha_1\beta_1\beta_2}(h_0;a,b,y)
\leq
\cases{
\kappa(a)+\kappa(b) + 2\epsilon, & \hbox{if $\alpha_1=\beta_1=0$};   \\
\log \mu_3 + \lambda(y) + 2\epsilon, & \hbox{if $\alpha_2=\beta_2=0$}; \\
\alpha_2 \kappa(a) + \alpha_1 \log \mu_3 
+ \beta_1 \lambda(y) + \beta_2 \kappa(b)+ 2\epsilon, & \hbox{otherwise}.
}
 \] 

\noindent{\textit{Proof of the claim:}}
Each $2$-star in figure \ref{fig56} is partitioned in $4$ (possibly 
empty) subwalks. Ignore intersections between these subwalks.  

We now construct bounds on the contributions of the arms of the $2$-star 
to the partition function.  Assume first that both arms intersect the
adsorbing surface.

In figure \ref{fig56} the tooth of the $2$-star is partitioned into 
two walks by the first vertex in the adsorbing surface.  The second 
part is an adsorbing walk of length $\lfloor \alpha_2 m_a\rfloor 
+ \sigma_a$, while the first part is a walk of length 
$\lfloor \alpha_1 m_a \rfloor$ from the trivalent node $i$ to the 
adsorbing surface.

The other arm of the $2$-star is similarly partitioned into an 
adsorbing walk from the trivalent node $i_0$ to the adsorbing 
surface, of length $\lfloor \beta_2 m_b \rfloor + \sigma_b$, and 
a walk from the trivalent node $i$ to the adsorbing surface, of 
length $\lfloor \beta_1 m_b\rfloor$ (see figure \ref{fig56}).

Denote the number of self-avoiding walks of length $n$ from a 
vertex at height $h_0$ to a vertex at height $h$ by $e_n(h_0,h)$. 
Note that the partition function introduced in section \ref{sec:2.1} 
for $x_3$-pulled (not necessarily positive) walks 
$C_n(y)=\sum_{h} e_n(0,h) y^{h}=\sum_{h} e_n(h_0,h) y^{h-h_0}$, 
independent of $h_0$.   Also, denote the number of positive 
self-avoiding walks from the origin making $v$ visits, including 
the origin, to the adsorbing surface by $c_n^+(v)$. Denote 
their partition function by $C_n^+(a)=\sum_v c_n^+(v)a^v = a\,C_n^+(a,1)$.

Then counting the walks in figure \ref{fig56} in terms of $e_n(h_0,h)$ and 
$c_n^+(v)$ and noting that $y\geq 1$ while $h_0\geq 0$, the following bound 
is obtained:
\begin{eqnarray}
& & \hspace{-3cm}
S_{m,m}^{\alpha_2\alpha_1\beta_1\beta_2}(h_0;a,b,y) \nonumber \\
& & \hspace{-2cm} 
\leq \left[ \sum_h e_{\lfloor\alpha_1 m \rfloor}(0,h)\, y^{h-h_0}\,
 e_{\lfloor\beta_1 m \rfloor}(0,h) \right]
\left[\sum_v c_{\lfloor \alpha_2 m \rfloor+\sigma_a}^+(v)\,a^v \right]
\left[\sum_v c_{\lfloor \beta_2 m \rfloor+\sigma_b}^+(v)\,b^v \right] \nonumber \\
&\leq& c_{\lfloor \alpha_1 m\rfloor} 
 \left[ \sum_h e_{\lfloor\beta_1 m \rfloor}(0,h)\, y^h \right]
\left[\sum_v c_{\lfloor \alpha_2 m \rfloor+\sigma_a}^+(v)\,a^v \right]
\left[\sum_v c_{\lfloor \beta_2 m \rfloor\sigma_b}^+(v)\,b^v \right] \nonumber \\
&\leq& c_{\lfloor \alpha_1 m\rfloor}
C_{\lfloor\beta_1 m \rfloor}(y)
C_{\lfloor \alpha_2 m \rfloor+\sigma_a}^+(a)\, 
C_{\lfloor \beta_2 m \rfloor+\sigma_b}^+(b) .
\label{a4}
\end{eqnarray}
Observe that the right hand side of 
equation (\ref{a4}) is independent of $h_0$.  

For the three cases when at least one arm of the $2$-star does not intersect the
adsorbing surface (here, $\alpha_2=0$ or $\beta_2=0$, or both), it can be verified that
the same final upper bound in equation (\ref{a4}) applies.

In addition,
\[ \lim_{n\to\infty} \frac{1}{n} \log C_n^+(a) = \kappa(a)\quad\cite{HTW}
\quad\hbox{and for}\quad y\geq 1\quad \lim_{n\to\infty} \frac{1}{n} \log C_n(y)
  = \lambda(y) \quad \cite{Rensburg2016a} .\]
Thus, for an arbitrary and fixed $\epsilon>0$ there exists an $M>0$ 
such that for all $n>M$, $\frac{1}{n} \log C_n^+(a) \leq \kappa(a) + \epsilon$,
and $\frac{1}{n} \log C_n(y) \leq \lambda(y) +\epsilon$.
Using these in equation (\ref{a4}) shows that there are large, but finite
values of $m>M$ (a function of $(h_0,a,b)$), such that
\begin{equation*}
\hspace{-2cm}
\frac{1}{m} \log S_{m,m}^{\alpha_2\alpha_1\beta_1\beta_2}(h_0;a,b,y)
\leq
\cases{
\kappa(a)+\kappa(b) + 2\epsilon, & \hbox{if $\alpha_1=\beta_1=0$};   \\
\lambda(y) + \log \mu_3 + 2\epsilon, & \hbox{if $\alpha_2=\beta_2=0$}; \\
\alpha_2 \kappa(a) + \alpha_1 \log \mu_3 
+ \beta_1 \lambda(y) + \beta_2 \kappa(b)+ 2\epsilon, & \hbox{otherwise}.
}
\end{equation*}
This completes the proof of the claim. $\square$
\vspace{2mm}

Simplify the cases in the claim by maximizing the expression
\begin{equation}
\alpha_2 \kappa(a) + \alpha_1 \log \mu_3 
+ \beta_1 \lambda(y) + \beta_2 \kappa(b)
\end{equation}
with $\alpha_1+\alpha_2=1$, $\beta_1+\beta_2=1$, and
$0\leq \alpha_1,\beta_1\leq 1$. Since $\alpha_1=0$ if and only if
$\beta_1=0$, this is a maximum if either $\alpha_1=\beta_1=0$,
or $\alpha_1=\beta_1=1$.  This shows that, for fixed values of $(h_0,a,b,y)$,
there exists an $M$ such that for all $m>M$,
\begin{equation}
\frac{1}{m} 
\log S_{m,m}^{\alpha_2\alpha_1\beta_1\beta_2}(h_0;a,b,y)
\leq \max ( \kappa(a)+\kappa(b), \lambda(y)+\log\mu_3) + 2\epsilon .
\end{equation}
Exponentiating the above and noting that the right hand side is independent
of $(\alpha_2,\alpha_1,\beta_1,\beta_2)$,
\begin{equation}
\max_{\alpha_1\beta_1}
S_{m,m}^{\alpha_2\alpha_1\beta_1\beta_2}(h_0;a,b,y)
\leq e^{m(\hbox{MAX}+2\epsilon)}
\label{a6}
\end{equation}
where
\[ \hbox{MAX}=\max ( \kappa(a)+\kappa(b), \lambda(y)+\log\mu_3) . \]

Proceed by taking logarithms of equation (\ref{eqn57}).  Divide by $(2t+1)\,m$, 
and take the limit superior as $m\to\infty$ on the left hand side.  Since 
$H=O(m)$, this gives
\begin{eqnarray}
& & \hspace{-2cm}
\limsup_{m\to\infty} \frac{1}{(2t+1)\,m} 
\log K^{(t)}(m,m,a,b,y) \nonumber \\ 
&\leq& \limsup_{m\to\infty}  \frac{t}{(2t+1)\,m} \log 
\left[ \max_{\alpha_1\beta_1}
S_{m,m}^{\alpha_2\alpha_1\beta_1\beta_2}(h_0;a,b,y)\right] \nonumber \\
& &+ \limsup_{m\to\infty} \frac{1}{(2t+1)\,m}
\log \left[ \max_{h_0} C_{m}(h_0;b,y) \right]  .
\label{a3}
\end{eqnarray}
By equation (\ref{a6}),
\begin{eqnarray}
& & \hspace{-2cm}
\limsup_{m\to\infty} \frac{1}{(2t+1)\,m} 
\log K^{(t)}(m,m,a,b,y) \nonumber \\ 
&\leq& \frac{t}{(2t+1)} \left( \hbox{MAX} + 2\epsilon \right) 
+ \limsup_{m\to\infty} \frac{1}{(2t+1)\,m}
\log \left[ \max_{h_0} C_{m}(h_0;b,y) \right]  .
\label{a7}
\end{eqnarray}
It remains to examine the last term above.  We notice that it 
cannot exceed a linear combination of $\kappa(b)$ and $\lambda(y)$.  Thus,
if $\hbox{MAX}=\lambda(y)+\log \mu_3$ then $\kappa(b) \leq \lambda(y)$
and the last term is bounded by $\lambda(y)$.  On the other hand,
if $\hbox{MAX}=\kappa(a)+\kappa(b)$, then either $\kappa(b)\leq \lambda(y)$
and the last term is bounded by $\lambda(y)$, or $\kappa(b)\geq \lambda(y)$
and the last term is bounded by $\kappa(b)$. Collecting these results and
taking $\epsilon\to 0^+$ gives
\begin{eqnarray*}
& & \hspace{-2cm}
\limsup_{m\to\infty} \frac{1}{(2t+1)\,m} 
\log K^{(t)}(m_a,m_b,a,b,y) \\
&\leq& \frac{1}{2t+1}\max\left(
t\kappa(a)+(t+1)\kappa(b),
t\kappa(a)+t\kappa(b)+\lambda(y),
(t+1) \lambda(y)+t\log\mu_3
\right) .
\end{eqnarray*}
This completes the case $y\geq 1$.
\vspace{2mm}

\noindent{{\bf The case $\mathbf{y < 1}$}:} 
If $y<1$, then we notice that $K^{(t)}(m_a,m_b,a,b,y) < K^{(t)}(m_a,m_b,a,b,1)$.
If $y=1$, then
\begin{eqnarray*}
& & \hspace{-2cm}
\limsup_{m\to\infty} \frac{1}{(2t+1)\,m} 
\log K^{(t)}(m,m,a,b,1) \\
&\leq& \frac{1}{2t+1}\max\left(
t\kappa(a)+(t+1)\kappa(b),
t\kappa(a)+t\kappa(b)+\log \mu_3,
(2t+1)\log\mu_3 \right) .
\end{eqnarray*}
Evaluating the maximum gives the upper bound
$(t\,\kappa(a)+(t+1)\kappa(b))/(2t+1)$. This completes the proof of 
the theorem. \qed

\begin{figure}[h!]
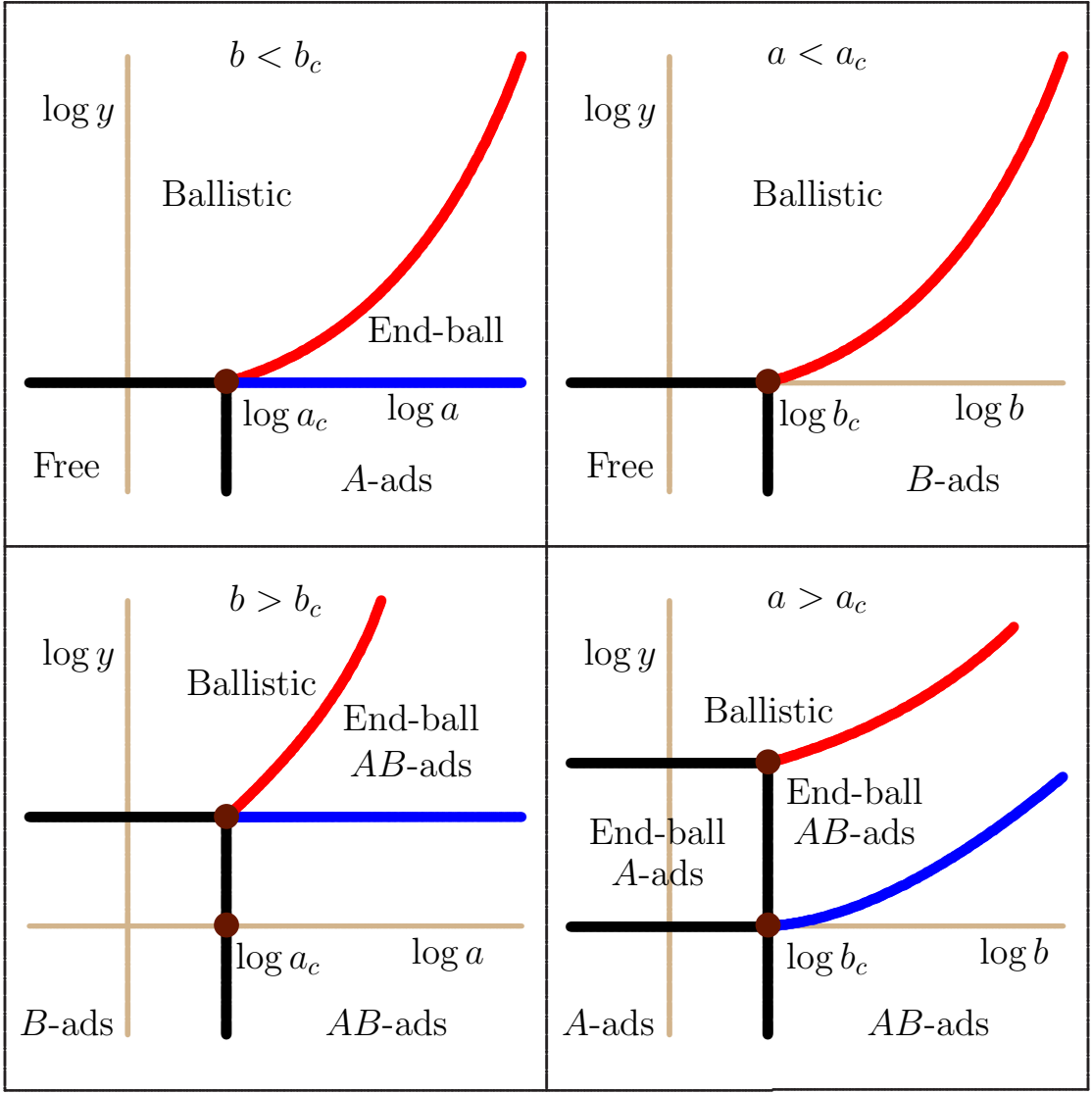

\beginpicture
\setcoordinatesystem units <0.95pt,1.05pt>
\setplotarea x from 0 to 400, y from -10 to 410
\setplotsymbol ({\scalebox{0.25}{$\bullet$}})
\plot -10 0 -10 400 430 400 430 0 -10 0 /
\plot 210 0 210 400 /  \plot -10 200 430 200 /
\setplotsymbol ({\scalebox{0.5}{$\bullet$}})
\color{Tan}
\plot 0 260 200 260 /  \plot 40 220 40 380 /
\setplotsymbol ({\scalebox{1.0}{$\bullet$}})
\setquadratic \color{Red} \plot 80 260 150 300 200 380 /
\setlinear \color{Black} \plot 0 260 80 260 80 220 /
\setlinear \color{Blue} \plot 80 260 200 260 /
\color{Sepia} \put {\scalebox{2.5}{$\bullet$}} at 80 260 
\color{Black}
\put {\Large$\log a$} at 160 250  \put {\Large$\log y$} at 20 360
\put {\Large$b<b_c$} at 100 380
\put {\Large$\log a_c$} at 104 248
\put {\Large Ballistic} at 80 330
\put {\Large Free} at 15 230
\put {\Large$A$-ads} at 145 225
\put {\Large End-ball} at 165 280
\setplotsymbol ({\scalebox{0.5}{$\bullet$}})
\color{Tan}
\plot 240 260 420 260 /  \plot 260 220 260 380 /
\setplotsymbol ({\scalebox{1.0}{$\bullet$}})
\setquadratic \color{Red} \plot 300 260 370 300 420 380 /
\setlinear \color{Black} \plot 220 260 300 260 300 220 /
\color{Sepia} \put {\scalebox{2.5}{$\bullet$}} at 300 260 
\color{Black}
\put {\Large$\log b$} at 390 250  \put {\Large$\log y$} at 240 360
\put {\Large$a<a_c$} at 320 380
\put {\Large$\log b_c$} at 321 248
\put {\Large Ballistic} at 320 330
\put {\Large Free} at 240 230
\put {\Large$B$-ads} at 375 225
\setplotsymbol ({\scalebox{0.5}{$\bullet$}})
\color{Tan}
\plot 220 60 420 60 /  \plot 260 20 260 180 /
\setplotsymbol ({\scalebox{1.0}{$\bullet$}})
\setquadratic \color{Blue} \plot 300 60 355 75 420 115 /
\setquadratic \color{Red} \plot 300 120 355 140 400 170 /
\setlinear \color{Black} \plot 220 60 300 60 / \plot 220 120 300 120 /
                         \plot 300 60 300 20 / \plot 300 60 300 120 /
\color{Sepia} \multiput {\scalebox{2.5}{$\bullet$}} at 300 60 300 120 / 
\color{Black}
\put {\Large$\log b$} at 400 50  \put {\Large$\log y$} at 240 160
\put {\Large$a>a_c$} at 320 180
\put {\Large$\log b_c$} at 324 48
\put {\Large Ballistic} at 300 140
\put {\Large$A$-ads} at 235 25
\put {\Large$AB$-ads} at 365 25
\put {\Large End-ball} at 255 95
\put {\Large$A$-ads} at 255 80
\put {\Large End-ball} at 335 110
\put {\Large $AB$-ads} at 335 95
\setplotsymbol ({\scalebox{0.5}{$\bullet$}})
\color{Tan}
\plot 0 60 200 60 /  \plot 40 20 40 180 /
\setplotsymbol ({\scalebox{1.0}{$\bullet$}})
\setquadratic \color{Blue} \plot 80 100 110 100 200 100 /
\setquadratic \color{Red} \plot 80 100 120 140 143 180 /
\setlinear \color{Black}  \plot  0 100  80 100  80 20 /
\color{Sepia} \multiput {\scalebox{2.5}{$\bullet$}} at 80 60 80 100 / 
\color{Black}
\put {\Large$\log a$} at 170 50  \put {\Large$\log y$} at 20 160
\put {\Large$b>b_c$} at 100 180
\put {\Large$\log a_c$} at 101 48
\put {\Large Ballistic} at 90 150
\put {\Large $AB$-ads} at 155 120
\put {\Large End-ball} at 155 137
\put {\Large$B$-ads} at 15 25
\put {\Large$AB$-ads} at 145 25

\color{black}
\normalcolor
\endpicture
\caption{Phase diagrams of pulled and adsorbing uniform combs
at constant $a$ or constant $b$.  The top left diagram is for
fixed $b<b_c$ and the top right for fixed $a<a_c$.  The top right
diagram is qualitatively similar to that of a pulled adsorbing
self-avoiding walk.  On the bottom left $b>b_c$ and the backbone
is adsorbed or partially adsorbed in all but the ballistic phase.
The bottom right phase diagram has five phases, similar to what
was established in pulled adsorbing copolymeric stars in reference
\cite[see figure 10(b)]{Rensburg2022}.}
\label{F5}
\end{figure}

\subsection{Phase boundaries of uniform combs}

\subsubsection{The case $b<b_c$:}
In this case $\kappa(b)=\log\mu_3$ and 
$$\zeta(a,b,y)=\max \left[\Sfrac{t}{2t+1}\kappa(a) + \Sfrac{t+1}{2t+1} 
\log\mu_3, \Sfrac{1}{2t+1}\lambda(y) + \Sfrac{t}{2t+1}(\kappa(a)+\log\mu_3),
\Sfrac{t+1}{2t+1} \lambda(y) + \Sfrac{t}{2t+1} \log \mu_3\right].$$
For $b<b_c$ and $y<1$, $\zeta(a,b,y)= \Sfrac{t}{2t+1}\kappa(a) 
+ \Sfrac{t+1}{2t+1} \log\mu_3$ which has a critical point at $a=a_c$ 
and there is a phase boundary from $A$-adsorbed to $A$-desorbed and 
no ballistic phases. 

For $y>1$, $\lambda(y)>\log\mu_3$ and $\kappa(a)\geq \log\mu_3$.  For 
$a<a_c$,  $\kappa(a)= \log\mu_3$ and the last two terms  
are always larger than the first. Thus for $b<b_c$, $y>1$, $a<a_c$, 
$\zeta(a,b,y)= \Sfrac{t+1}{2t+1} \lambda(y) + \Sfrac{t}{2t+1} \log \mu_3$ 
which is a fully ballistic phase. For $a>a_c$,  $\kappa(a)> \log\mu_3$ 
and we need to compare $\lambda(y)$ and $\kappa(a)$.  If 
$\lambda(y)<\kappa(a)$, then the middle term is larger than both the first and 
the last term since $\lambda(y)>\log\mu_3$. Hence 
$\zeta(a,b,y)= \Sfrac{1}{2t+1}\lambda(y) 
+ \Sfrac{t}{2t+1}(\kappa(a)+\log\mu_3)$ 
and we  have an end-ballistic/$A$-adsorbed phase. For  $\lambda(y)>\kappa(a)$, 
then the last term is larger than both the first term and the middle term since 
$\lambda(y)>\log\mu_3$. Hence $\zeta(a,b,y)= \Sfrac{t+1}{2t+1} \lambda(y) 
+ \Sfrac{t}{2t+1} \log \mu_3$ and we  have a fully ballistic phase.

It follows that for $b<b_c$ and $y<1$, there is a phase boundary from 
a free phase to an $A$-adsorbed phase at $a=a_c$ (this phase boundary 
is illustrated in figure~\ref{F5} (top left)).  If $b<b_c$ and $y>1$, 
there is a phase boundary at $\lambda(y)=\kappa(a)$ (or 
$y=\lambda^{-1}(\kappa(a))$) which divides a fully ballistic and 
an end-ballistic (and $A$-adsorbed) phase (the case $a<a_c$ 
lies in the fully ballistic phase). Along $y=1$ and $a>a_c$ runs a
phase boundary separating the $A$-adsorbed and end-ballistic phases.  
Thus the boundary $y=1$ bounds four different phases 
(free, ballistic, $A$-adsorbed and end-ballistic); 
see figure~\ref{F5} (top left).

The phase boundary between the ballistic and end-ballistic phases
is first order, while the other phase boundaries are presumably 
continuous.  Since $\lambda(y)$ is asymptotic to $\log y$
\cite{Rensburg2013} and $\kappa (a)$ is asymptotic to 
$\log a + \log \mu_2$ \cite{Rensburg2013,Rychlewski}, 
the phase boundary between the ballistic and end-ballistic phases 
is asymptotic to $\log y = \log a + \log \mu_2$.

\subsubsection{The case $a<a_c$:} 
If $a<a_c$ then $\kappa(a)=\log\mu_3$ and 
$$\zeta(a,b,y)=\max \left[\Sfrac{t}{2t+1}\log\mu_3 
+ \Sfrac{t+1}{2t+1}\kappa(b), \Sfrac{1}{2t+1}\lambda(y) 
+ \Sfrac{t}{2t+1}(\kappa(b)+\log\mu_3),
\Sfrac{t+1}{2t+1} \lambda(y) + \Sfrac{t}{2t+1} \log \mu_3 \right].$$
The phase diagram of this case is illustrated in figure \ref{F5} (top right).

For $y<1$, $\lambda(y)=\log\mu_3$ and $\kappa(b)\geq \log\mu_3$ 
thus the first term is always greater than the other two. Thus for 
$a<a_c$ and $y<1$, the free energy $\zeta(a,b,y) 
= \Sfrac{t+1}{2t+1}\kappa(b) + \Sfrac{t}{2t+1} \log\mu_3$ 
which has a critical point at $b=b_c$ and there is a phase boundary from 
$B$-adsorbed to $B$-desorbed and no ballistic phases.

For $y>1$, $\lambda(y)>\log\mu_3$ and $\kappa(b)\geq \log\mu_3$.  
For $b<b_c$,  $\kappa(b)= \log\mu_3$ and the last term is always 
larger than the other two. Thus for $a<a_c$, $y>1$, $b<b_c$, the free energy
$\zeta(a,b,y)= \Sfrac{t+1}{2t+1} \lambda(y) + \Sfrac{t}{2t+1} \log \mu_3$
which is a fully ballistic phase. For $b>b_c$,  $\kappa(b)> \log\mu_3$ 
and we need to compare $\lambda(y)$ and $\kappa(b)$.  If 
$\lambda(y)<\kappa(b)$, then the first term is greater than  the other two 
since $\lambda(y)>\log\mu_3$. Hence 
$\zeta(a,b,y)= \Sfrac{t}{2t+1}\log\mu_3 + \Sfrac{t+1}{2t+1}\kappa(b)$ 
which corresponds to a $B$-adsorbed/tooth-free phase (not end ballistic). 
For  $\lambda(y)>\kappa(b)$, then the last term is greater than both the first term
and the middle term since $\lambda(y)>\log\mu_3$. Hence 
$\zeta(a,b,y)= \Sfrac{t+1}{2t+1} \lambda(y) + \Sfrac{t}{2t+1} \log \mu_3$ 
and we  have a fully ballistic phase.

Thus for $a<a_c$ and $y<1$, there is a phase boundary from a 
free phase to a $B$-adsorbed phase at $b=b_c$. 

For $y>1$, there is a phase boundary at 
$\lambda(y)=\kappa(b)$ ($y=\lambda^{-1}(\kappa(b))$) which divides 
a ballistic phase for large $y$ from the $B$-adsorbed phase at large
$b$ (the region $b<b_c$ and $y>1$ is in the fully ballistic phase).  
There is no end-ballistic phase in this phase diagram. At the point
$b=b_c$ and $y=1$ there are three phase boundaries meeting, as shown
in figure~\ref{F5} (top right).

The phase boundary between the ballistic and $B$-adsorbed phases is
first order, while the other phase boundaries are presumably continuous.  
The phase boundary between the ballistic and $B$-adsorbed phases is asymptotic 
to $\log y = \log b + \log \mu_2$.

\subsubsection{The case $b>b_c$:}
The phase diagram is illustrated in figure \ref{F5} (bottom left).
If $b>b_c$ then $\kappa(b)>\log\mu_3$.  In the case that $a<a_c$, 
then, as discussed above, for $y<1$ or $y>1$ and $\lambda(y)<\kappa(b)$,  
the free energy $\zeta(a,b,y)= \Sfrac{t+1}{2t+1}\kappa(b) 
+ \Sfrac{t}{2t+1} \log\mu_3$ which corresponds to a 
$B$-adsorbed/tooth-free phase, while for $y>1$ and  
$\lambda(y)>\kappa(b)$, the free energy $\zeta(a,b,y) 
= \Sfrac{t+1}{2t+1} \lambda(y) + \Sfrac{t}{2t+1} \log \mu_3$ 
and we  have a fully ballistic phase.

If $a>a_c$, $\kappa(a)>\log\mu_3$, and if $y<1$, then 
$\lambda(y)=\log\mu_3$ and the first term is larger than the other two so that 
$\zeta(a,b,y)=\Sfrac{t}{2t+1}\kappa(a) + \Sfrac{t+1}{2t+1} \kappa(b)$ 
which corresponds to an $AB$-adsorbed phase.  If $y>1$,  we need to 
compare all three terms.  We get three potential boundaries, namely:
(1) $\lambda(y)=\kappa(b)$ (when the first two terms are equal; these terms 
are larger than the third term since $\kappa(a)>\log\mu_3$);
(2) $\lambda(y)=\Sfrac{t}{t+1}(\kappa(a)-\log\mu_3)+ \kappa(b)$ 
(when the first and last terms are equal, the middle term is the maximum);
(3) lastly, $\lambda(y)=(\kappa(a)-\log\mu_3)+ \kappa(b)$ (the last 
two terms are equal; these terms are larger than the first term since 
$\lambda(y)>\kappa(b)$). 

Thus, when $a>a_c$ and $y>1$ there are three regions.  These are
$\lambda(y)<\kappa(b)$ where the free energy is given by the 
first term and we have an $AB$-adsorbed phase; $\kappa(b)<\lambda(y)
<(\kappa(a)-\log\mu_3)+ \kappa(b)$ where the free energy is given 
by the middle term and we have an end-ballistic phase; 
$\lambda(y)>(\kappa(a)-\log\mu_3)+ \kappa(b)$ where the free energy 
is given by the last term and we have a fully ballistic phase. 

There are four phases (namely $B$-adsorbed, $AB$-adsorbed, ballistic
and end-ballistic) meeting at a multicritical point located at
$\lambda(y)=\kappa(b)$ and $a=a_c$. The phase boundary
$y=\lambda^{-1}(\kappa(b))$ separates the $B$-adsorbed and 
ballistic phases when $a<a_c$, and the $AB$-adsorbed and end-ballistic
phases when $a>a_c$.  The phase boundary between the $B$-adsorbed and
$AB$-adsorbed phases is $a=a_c$, and the phase boundary between the
ballistic and end-ballistic phases is 
$\lambda(y)=(\kappa(a)-\log\mu_3)+ \kappa(b)$.

The phase boundary between the ballistic and end-ballistic phases
is first order, but the other phase boundaries are presumably
continuous.  The phase boundary between the ballistic and end-ballistic 
phases is asymptotic to $\log y = \log a + \log b + 
2 \log \mu_2 - \log \mu_3$.

\subsubsection{The case $a>a_c$:}
The phase diagram is illustrated in figure \ref{F5} (bottom right).
Since $a>a_c$, $\kappa(a)>\log\mu_3$. 

If $b<b_c$ and $y<1$ then $\zeta(a,b,y)= \Sfrac{t}{2t+1}\kappa(a) 
+ \Sfrac{t+1}{2t+1} \log\mu_3$ which corresponds to an $A$-adsorbed
phase.  

If $b<b_c$ and $y>1$, then for $\lambda(y)<\kappa(a)$, the free
energy $\zeta(a,b,y)= \Sfrac{1}{2t+1}\lambda(y) 
+ \Sfrac{t}{2t+1}(\kappa(a)+\log\mu_3)$ and we  have an 
end-ballistic/$A$-adsorbed phase. For $\lambda(y)>\kappa(a)$,  
the free energy $\zeta(a,b,y) = \Sfrac{t+1}{2t+1} \lambda(y) 
+ \Sfrac{t}{2t+1} \log \mu_3$ and we have a ballistic phase. 

These phases for $b<b_c$ are shown as ``$A$-adsorbed", ``end-ballistic", 
and ``ballistic" in figure \ref{F5} (bottom right).

The case $b>b_c$ and $y<1$ gives $\zeta(a,b,y)=\Sfrac{t}{2t+1}\kappa(a) 
+ \Sfrac{t+1}{2t+1} \kappa(b)$ which corresponds to an $AB$-adsorbed phase.  

If $y>1$ and $b>b_c$,  we have three phases: 
(1) If $\lambda(y)<\kappa(b)$ then the free energy is given by the 
first term and we have an $AB$-adsorbed phase; 
(2) if $\kappa(b)<\lambda(y)<(\kappa(a)-\log\mu_3) + \kappa(b)$ 
then the free energy is given by the middle term and 
we have an end-ballistic/$AB$-adsorbed phase; (3) if 
$\lambda(y)>(\kappa(a)-\log\mu_3) + \kappa(b)$ where the free energy 
is given by the last term and we have a ballistic phase.  

These phases for $b>b_c$ are shown as ``ballistic", 
``end-ballistic/$AB$-adsorbed", and ``$AB$-adsorbed" in 
figure \ref{F5} (bottom right).  

The phase boundary between the ballistic and end-ballistic phases 
is asymptotic to $\log y = \log b + \log a +2 \log \mu_2 - \log \mu_3$,
and the phase boundary between the end-ballistic and $AB$-adsorbed 
phases is asymptotic to $\log y = \log b + \log \mu_2$. Notice that 
these curves are parallel in the $(\log y)$-$(\log b)$ plane.

\subsubsection{The $(a,b)$-plane for $y > 1$:}
\begin{figure}[!ht]
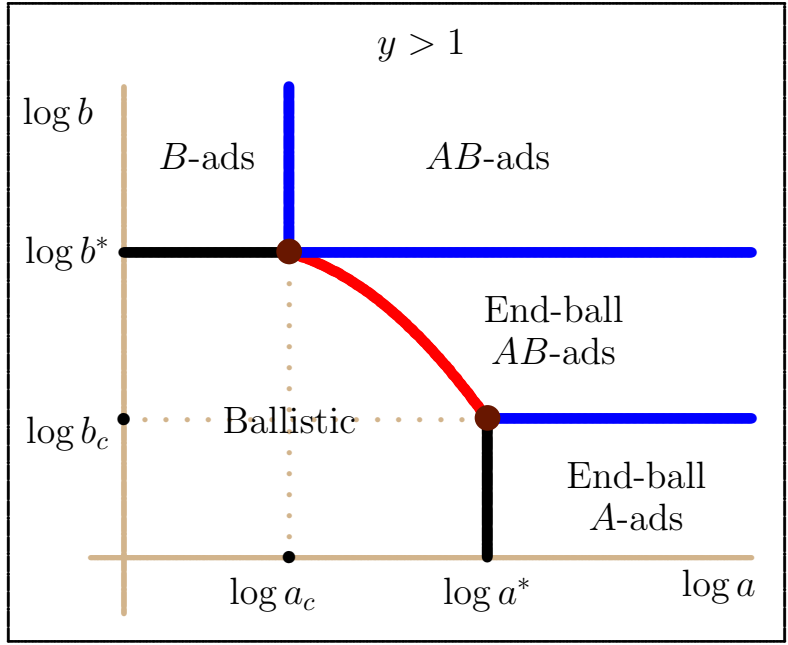

\beginpicture
\setcoordinatesystem units <1.25pt,1.05pt>
\setplotarea x from -80 to 210, y from -30 to 230
\setplotsymbol ({\scalebox{0.25}{$\bullet$}})
\plot -25 -20 -25 210 210 210 210 -20 -25 -20 /
\setplotsymbol ({\scalebox{0.5}{$\bullet$}})
\color{Tan}
\plot 0 10 200 10 /  \plot 10 -10 10 180 /
\setdots <8pt>
\plot 60 10 60 180 /  \plot 10 60 200 60 /
\setsolid
\setplotsymbol ({\scalebox{1.0}{$\bullet$}})
\setquadratic \color{Red} \plot 120 60 90 100 60 120  /
\setlinear \color{Black} \plot 120 10 120 60 / \plot  10 120 60 120 /
\setlinear \color{Blue} \plot 120 60 200 60 /  \plot 60 180 60 120  200 120 /
\color{Sepia} \multiput {\scalebox{2.5}{$\bullet$}} at 120 60 60 120 / 
\color{Black}
\put {\Large$\log a$} at 190 0  \put {\Large$\log b$} at -10 170
\put {\Large$y>1$} at 100 195
\put {\Large $AB$-ads} at 120 155
\put {\Large $B$-ads} at 35 155
\put {\Large End-ball} at 140 100
\put {\Large $AB$-ads} at   140 85
\put {\Large Ballistic} at 60 60 
\put {\Large End-ball} at 165 40
\put {\Large $A$-ads} at 165 25
\put {\large$\bullet$} at 60 10 \put {\Large$\log a_c$} at 55 -3
\put {\large$\bullet$} at 10 60 \put {\Large$\log b_c$} at -7 55
\put {\Large $\log b^*$} at -7 120
\put {\Large $\log a^*$} at 120 -3
\normalcolor\color{black}
\endpicture
\caption{Phase diagram in the $(\log a, \log b)$-plane for fixed $y>1$. 
Here, $a^* = \kappa^{-1}(\lambda(y))$ and $b^*=\kappa^{-1}(\lambda(y))$.
The curved phase boundary is given by the solution of 
$\kappa(b)=\lambda(y)-\kappa(a)+\log\mu_3$ and is concave down.}
\label{fig:abplane_ygt1}
\end{figure}

In figure \ref{fig:abplane_ygt1} we show a slice through the 
three dimensional phase diagram at fixed $y>1$.  At small values of 
$a$ and $b$ the system is ballistic.  If $b$ increases at fixed 
$a < a_c$ we enter a phase where the backbone is adsorbed for 
$b > b^*$, where $b^*= \kappa^{-1}(\lambda(y))$.  (The inverse 
function exists since $\kappa(b)$ is continuous and strictly monotone 
for $b>b_c$ \cite{HTW}.)  If $a$ increases at fixed $b<b_c$ the 
teeth adsorb when $a>a^*$ where $a^*=\kappa^{-1}(\lambda(y))$.
In this phase the last branch of the backbone is ballistic and 
the teeth are adsorbed.  If $a>a_c$ and $b>b^*$ the backbone and 
the teeth are adsorbed and we have an $AB$-adsorbed phase.  There 
is also a curved phase boundary between the ballistic phase and 
a phase in which the teeth are adsorbed, all except the last branch 
of the backbone are adsorbed, and the last branch is ballistic.  
This boundary is given by the solution of the equation 
$\lambda(y) + \log \mu_3 = \kappa(a) + \kappa(b)$, and the boundary 
is concave down in the $(\log a, \log b)$-plane.  To see this, 
take two points on the phase boundary, say $(\log a_1, \log b_1)$ and 
$(\log a_2, \log b_2)$.  The free energy is equal to 
$((t+1)\lambda(y)+ t \log \mu_3)/(2t+1)$ at both these points.  
Since the free energy is a convex function,  the free energy is 
less than or equal to $((t+1)\lambda(y)+ t \log \mu_3)/(2t+1)$ 
at the mid-point of the chord joining these points, so that 
this point is on or below the phase boundary.  The concavity 
of the boundary then follows from the mid-point theorem.

\section{Non-uniform combs}
\label{sec:nonuniform}

In this section we examine the limits that the teeth are short 
compared to the backbone, or the backbone is short compared to 
the teeth.

\subsection{Short teeth and long backbone}

As before, the number of teeth is fixed at $t$
and we are considering combs with initial vertex at the origin, in 
the half-space $x_3 \ge 0$, with $m_a$ edges in each tooth,
$m_b$ edges in each segment of the backbone, $v_A$ $A$-visits, 
$v_B$ $B$-visits and having the terminal vertex at height $h$, 
and with  partition function:
\begin{equation}
K^{(t)}(m_a,m_b,a,b,y) = \sum_{v_A,v_B,h}k^{(t)}(m_a,m_b,v_A,v_B,h)
\,a^{v_A}b^{v_B} y^h.
\label{eqn:generalPF}
\end{equation}
With $t$ and $m_a$ fixed, we have the following lemma as $m_b\to\infty$:

\begin{figure}[h!]
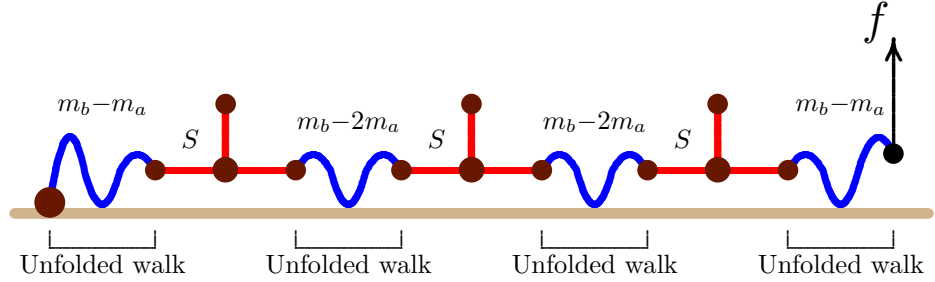

\beginpicture
\setcoordinatesystem units <1.33pt,1.25pt> 
\setplotarea x from -80 to 200, y from -25 to 60

\setplotsymbol ({\scalebox{1.00}{$\bullet$}})
\color{Tan}
\plot -20 -3 240 -3 /

\setplotsymbol ({\scalebox{0.16}{$\bullet$}})
\color{Black}
\setsolid
\plot -10 -8 -10 -13 20 -13 20 -8 /
\plot 60 -8 60 -13 90 -13 90 -8 /
\plot 130 -8 130 -13 160 -13 160 -8 /
\plot 200 -8 200 -13 230 -13 230 -8 /
\multiput {Unfolded walk} at 5 -18 75 -18 145 -18 215 -18 /
\multiput {$m_b{-}m_a$} at 5 30 215 30 /
\multiput {$m_b{-}2m_a$} at 75 25 145 25 /

\setplotsymbol ({\scalebox{0.67}{$\bullet$}})
\color{Red}
\plot 20 10 60 10 /  \plot 40 10 40 30 /
\plot 90 10 130 10 / \plot 110 10 110 30 /
\plot 160 10 200 10 /  \plot 180 10 180 30 /

\setquadratic
\color{Blue}
\plot -10 0 -5 20 0 10 5 0 10 10 15 15 20 10 /
\plot 60 10 65 15 70 10 75 0 80 10 85 15 90 10 /
\plot 130 10 135 15 140 10 145 0 150 10 155 15 160 10 /
\plot 200 10 205 15 210 10 215 0 220 10 225 20 230 15 /
\setlinear

\color{Sepia}
\multiput {\scalebox{2.00}{$\bullet$}} at 20 10 90 10 160 10
60 10 130 10 200 10 40 30 110 30 180 30 /
\multiput {\scalebox{2.50}{$\bullet$}} at 40 10 110 10 180 10 /
\multiput {\scalebox{3.00}{$\bullet$}} at -10 0 /
\setplotsymbol ({\scalebox{0.2}{$\bullet$}})

\setplotsymbol ({\scalebox{0.3}{$\bullet$}})
\color{black}
\multiput {$S$} at 30 20 100 20 170 20 /
\arrow <8pt> [.2,.67] from 230 15 to 230 50
\put {\scalebox{2}{$\bullet$}} at 230 15 
\put {\scalebox{1.75}{$f$}} at 225 55

\normalcolor
\endpicture
\caption{Concatenating unfolded walks with the structure $S$
recursively to form a pulled adsorbing comb.  The branches 
along the backbone each have length $m_b$ and the straight 
teeth are each of length $m_a$.  Each of the three branches 
of the structure $S$ has length $m_a$.}
\label{fig11}
\end{figure}

\begin{lemm}
If $t$ and $m_a$ are both fixed, with the total number of edges in 
the comb being given by $n=(t+1)\,m_b + t\,m_a$,  the free energy is given by
$$\lim_{m_b\to\infty} \Sfrac{1}{(t+1)\,m_b + t\,m_a} \log K^{(t)}(m_a,m_b,a,b,y) 
=\max[\kappa(b), \lambda(y)] =\psi(b,y).$$
\label{lem:teethfixed}
\end{lemm}
\Pr 
Define a structure $S$ to be $2m_a$ edges in the $x_1$-direction and 
$m_a$ edges starting at the centre vertex of this subwalk and extending 
in the $x_3$-direction.  This is a $3$-star with $m_a$ edges in each arm, 
and each arm being a straight line.  Construct a comb with $t$
teeth, each tooth with $m_a$ edges, and a backbone of total 
length $(t+1)\,m_b$, as follows (see figure \ref{fig11}).  
Take a walk unfolded in the $x_1$-direction of length $m_b-m_a$, 
concatenated with a structure $S$, then recursively
concatenated with a walk unfolded in the $x_1$-direction of length 
$m_b-2m_a$, concatenated with an $S$, a total of $t$ times, and ending 
with a walk unfolded in the $x_1$-direction of length $m_b-m_a$.  

Consider the subset of these combs with the first $s$ nodes of 
degree $3$ in $x_3=0$ plane (that is, the first $s$ $x_1$-unfolded 
walks are loops) and the remaining $t-s$ nodes of degree $3$ located 
in the half-space $x_3 > 0$. If $s=0$ this gives the lower bound
$$\log K^{(t)}(m_a,m_b,a,b,y) 
\ge (m_b-m_a) \psi(b,y) + ((t-1)(m_b-2m_a) \lambda(y) 
+ (m_b-m_a)\lambda(y) + o(m_b).$$
If $s=t$ then the lower bound
$$\log K^{(t)}(m_a,m_b,a,b,y) 
\ge (m_b - m_a) \kappa(b) + 2t m_a \log b + (t-1) (m_b-2m_a) \kappa(b) 
+ (m_b-m_a) \psi(b,y) + o(m_b)$$
is obtained instead.  Also, if $0<s<t$ we have the lower bound
\begin{eqnarray*}
\log K^{(t)}(m_a,m_b,a,b,y) 
&\ge& (m_b-m_a) \kappa(b) + 2s m_a \log b + (s-1) (m_b-2m_a) \kappa(b) \\
&+& (m_b-2m_a)\psi(b,y) + (t-s-1)(m_b-2m_a)\lambda(y) \\
&+& (m_b-m_a)\lambda(y) + o(m_b).
\end{eqnarray*}
Since $s$ is arbitrary we can choose its value to optimize the bound.  
When $\kappa(b) > \lambda(y)$ the bound is most effective when 
$s=t$ and when $\lambda(y) > \kappa(b)$ it is most effective when 
$s=0$.  Dividing by $(t+1)\,m_b+t\,m_a$ and letting $m_b \to \infty$ 
with $t$ and $m_a$ fixed gives 
$$\liminf_{m_b\to\infty} 
\Sfrac{1}{(t+1)\,m_b + t\,m_a} \log K^{(t)}(m_a,m_b,a,b,y) \ge \max[\kappa(b),
\lambda(y)] =\psi(b,y).$$
To obtain an upper bound we can regard the comb as being composed 
of a backbone with $(t+1)\,m_b$ edges and $t$ teeth, and treat these 
components as being independent.  This gives the bound 
$$\log K^{(t)}(m_a,m_b,a,b,y) 
\le (t+1)\,m_b \psi(b,y) + t m_a \kappa(a) + o(m_b).$$
Dividing by $(t+1)\,m_b+t\,m_a$ and letting $m_b \to \infty$ with $t$ 
and $m_a$ fixed gives 
$$\limsup_{m_b\to\infty} \Sfrac{1}{(t+1)\,m_b + t\,m_a} 
\log K^{(t)}(m_a,m_b,a,b,y) \le \psi(b,y) .$$
This completes the proof.  \qed

This can be extended to the situation where $t$ is fixed and 
$m_b$ goes to infinity with $m_a=o(m_b)$.  We give this result 
in the next theorem.

\begin{theo}
If $t$ is fixed and $m_b$ goes to infinity with $m_a=o(m_b)$ 
the free energy is given by
$$\lim_{m_b\to\infty,m_a=o(m_b)} 
\Sfrac{1}{(t+1)\,m_b+t\,m_a} \log K^{(t)}(m_a,m_b,a,b,y) 
= \max[\kappa(b), \lambda(y)] =\psi(b,y).$$
\label{theo:teethslow}
\end{theo}
\Pr 
Since $m_a$ being fixed is a special case of $m_a=o(m_b)$ the result 
of the previous lemma gives the lower bound
$$\liminf_{{m_b\to\infty},{ m_a=o(m_b)}} 
\Sfrac{1}{(t+1)\,m_b+t\,m_a} \log K^{(t)}(m_a,m_b,a,b,y) \ge \psi(b,y).$$
We can obtain an upper bound by considering the case where the 
backbone and teeth behave independently. Again we have 
$$\log K^{(t)}(m_a,m_b,a,b,y) \leq (t+1)\,m_b \psi(b,y) + t m_a \kappa(a) 
+ o(m_b).$$  
Dividing by $(t+1)\,m_b + t\,m_a$ and letting $m_b \to \infty$ 
with $t$ fixed and $m_a = o(m_b)$ gives 
$$\limsup_{{m_b\to\infty},{ m_a=o(m_b)}} 
\Sfrac{1}{(t+1)\,m_b+t\,m_a} \log K^{(t)}(m_a,m_b,a,b,y) \le \psi(b,y),$$
which completes the proof.
\qed

This result is expected, since the backbone dominates the
partition function in the limit;  this is the walk limit in this model.

\subsection{Long teeth and short backbone}

Assume that $t$ is fixed, $m_b=o(m_a)$ and that if the limit
$m_a \to \infty$ is taken, then $m_b\to\infty$ as well.  
In this case the teeth are long while
the backbone is short, and the partition function is again given by
equation \BRef{eqn:generalPF}.  An upper bound is found by
considering the teeth to be self-avoiding walks independent of 
one another and of the backbone.  Denoting the
partition function of pulled adsorbing positive walks by
$C_n^+(b,y)$ (see equation \BRef{eqn3}), and the partition function
of adsorbing self-avoiding walks by $C_n(a)$, this shows that
\begin{eqnarray}
& \limsup_{m_a\to\infty} \Sfrac{1}{t\,m_a+(t+1)\,o(m_a)}
 \log K^{(t)}(m_a,o(m_a),a,b,y) \cr 
& \leq  
 \limsup_{m_a\to\infty} \Sfrac{1}{t\,m_a+(t+1)\,o(m_a)}
 \left( (t+1)\log C_{o(m_a)}^+(b,y) + t \log C_{m_a}(a)\right)
= \kappa(a) .
\label{eqn11a}
\end{eqnarray}

For a lower bound,  equation (\ref{eq:15}) holds here and gives 

\begin{eqnarray}
  K^{(t)}(m_a,m_b,a,b,y) \geq 
\left[L_{o(m_a)}^{\dagger[1]}({>}w,b)\right]^{t} \;
\left[L_{o(m_a)}^{\dagger[1]}(b)\right]^{1} \;
\left[ C_{m_a-u}^{(w)}(a)\right]^t .  
\label{eq:15oma}
\end{eqnarray}

Take logarithms, divide by $t\,m_a + (t+1)\,o(m_a)$, and then
take $m_a\to\infty$.  By lemma \ref{T1}, 

\begin{equation}
\liminf_{m_a\to\infty} \Sfrac{1}{t\,m_a + (t+1)\,o(m_a)} \log
K^{(t)}(m_a,o(m_a),a,b,y) \geq \kappa^{(w)}(a) .
\end{equation}

Since $w$ is arbitrary, one may take $w\to\infty$ on the 
right hand side.  By lemma \ref{T1}, $\kappa^{(w)}(a)\to \kappa(a)$,
and comparing this result to equation \BRef{eqn11a}, the following
theorem is obtained.

\begin{theo}
 $\displaystyle \lim_{m_a\to\infty} \Sfrac{1}{t\,m_a+(t+1)\,o(m_a)} 
 \log K^{(t)}(m_a,o(m_a),a,b,y) = \kappa(a)$. \hfill \qed
 \label{theo:sec4star}
\end{theo}

This result is expected, since the adsorbing teeth dominate the
partition function in the limit.  Since the backbone has length
$o(m_a)$, the pulling force cannot pull any teeth from the adsorbing
plane, and if the teeth are adsorbed, then so is the backbone. 
This is the star limit in this model.

\section{The limit $t$ approaches infinity}
\label{sec:infinitet}

In this section the limit as $t\to\infty$ is examined.  In this limit
the comb becomes of infinite length, but with the lengths of teeth and
of the segments between teeth, fixed at $m_a$ and $m_b$ respectively.

Denote by $g(m_a,m_b,t)$ the number of combs from the origin, in the
bulk lattice ${\mathbb Z}^3$, with $t$ teeth of length $m_a$ and $t+1$ 
backbone segments of length $m_b$.  Then a comb from the origin of 
backbone length equal to $(s+t)\,m_b$ with $s+t-1$ teeth of length $m_a$,
can be cut, at the node along the backbone where the $s$-th tooth is attached,
into two subcombs with $s-1$ teeth and $t-1$ teeth respectively, and a
tooth which is a self-avoiding walk of length $m_a$.  This shows that
\begin{equation}
g (m_a,m_b,s+t-1) 
\leq c_{m_a}\, g(m_a,m_b,s-1)\,g (m_a,m_b,t-1),
\end{equation}
where $c_{m_a}$ accounts for the conformations of the orphaned tooth and
is the number of self-avoiding walks of length $m_a$ from the
origin in the lattice.  This shows that $c_{m_a}\,g(m_a,m_b,t)$ 
satisfies a submultiplicative inequality, and by reference \cite{Hille57} 
the connective constant \cite{Hammersley1957} of lattice combs is defined by
\begin{equation}
\zeta_{m_a,m_b} 
= \lim_{t\to\infty} \Sfrac{1}{t} \log g(m_a,m_b,t)
= \inf_{t\geq 0} \Sfrac{1}{t} \log \left( c_{m_a}\, g(m_a,m_b,t) \right) .
\end{equation}
Notice that $\zeta_{0,m_b} = m_b\log \mu_3$ where $\mu_3$ is the growth 
constant of the self-avoiding walk.

\subsection{Grafted combs pulled at an endpoint}

In this section we consider models of adsorbing combs in the half-lattice
where, as above, the backbone of the comb is taken to infinity by letting
$t\to\infty$ (but with $(m_a,m_b)$ fixed).  

For a lower bound, equation (\ref{eq:18}) holds here for $y\geq 1$, 
that is for $w<h_y$ and $a,b \geq 0$:

\begin{eqnarray}
  K^{(t)}(m_a,m_b,a,b,y) \geq 
K^{(t)} (m_a,m_b,0,0,y) \geq 
\left[\ell_{m_b}^{\dagger[3]}(h_y)\,y^{h_y}\right]^{t+1} \;
\left[ c_{m_a-u}^{\dagger[1],(w)}\right]^t ,
\label{eq:18repeated}
\end{eqnarray} 
where $K^{(t)} (m_a,m_b,0,0,y) $ denotes the partition function for 
combs with the backbone and the teeth  disjoint from the adsorbing 
surface (except at the origin).

Take logarithms, divide by $(t+1)\,m_b+t\,m_a$ and let $t\to\infty$:
\begin{eqnarray}
\fl \frac{1}{m_b+m_a} \log \left[\ell_{m_b}^{\dagger[3]} (h_y)\,y^{h_y} \right] 
&+ \frac{1}{m_b+m_a} \log c_{m_a-u}^{\dagger[1],(w)}
\label{eqn11}  \\
&\hspace{-2cm}
\leq \liminf_{t\to\infty} \frac{1}{(t+1)\,m_b+t\,m_a}
\log K^{(t)} (m_a,m_b,0,0,y) &= \Lambda_{inf}(m_a,m_b,0,0,y). \nonumber
\end{eqnarray}

To find an upper bound, use a construction similar to figure \ref{figg3}. 
Replace the backbone with a positive self-avoiding walk of length $(t+1)\,m_b$ 
from the origin, and pulled at its endpoint.  The teeth are
replaced by self-avoiding walks, each of length $m_a-u$.  This gives
\begin{equation}
 K^{(t)}(m_a,m_b,1,1,y)  \leq C_{(t+1)\,m_b}^+ (1,y)\, [c_{m_a-u}]^t . 
\end{equation}
Take logarithms, divide by $(t+1)\,m_b+t\,m_a$ and let $t\to\infty$:
\begin{eqnarray}
\fl \Lambda_{sup}(m_a,m_b,1,1,y) = 
\limsup_{t\to\infty} \frac{1}{(t+1)\,m_b 
+ t\,m_a} & \log  K^{(t)}(m_a,m_b,1,1,y) \nonumber \\
& \leq \frac{m_b}{m_b + m_a} \, \lambda(y) + \frac{1}{m_b+m_a} \log c_{m_a-u} .
\label{eqn13}
\end{eqnarray}
The bounds in equations \BRef{eqn11} and \BRef{eqn13} are valid in the square 
and in the cubic lattices.

The bounds above are independent of $a$ and $b$, but may be generalised by 
noting that for $0\leq a,b \leq 1$ and $y\geq 1$,
\begin{equation}
 K^{(t)}(m_a,m_b,0,0,y) \leq K^{(t)}(m_a,m_b,a,b,y) 
 \leq K^{(t)}(m_a,m_b,1,1,y) .
\end{equation}

\subsection{Adsorbed pulled combs}

In this section consider a fully adsorbed comb grafted at its first vertex, and 
being pulled at its endpoint by a vertical force. 

For a lower bound,  equation (\ref{eq:15}) holds here and gives 

\begin{eqnarray}
 K^{(t)}(m_a,m_b,a,b,y) \geq 
\left[L_{m_b}^{\dagger[1]}({>}w,b)\right]^{t} \;
\left[L_{m_b}^{\dagger[1]}(b)\right]^{1} \;
\left[ C_{m_a-u}^{(w)}(a)\right]^t .  
\label{eq:15repeat}
\end{eqnarray}

Taking logarithms, dividing by $(t+1)\,m_b+t\,m_a$ and then taking 
$t\to\infty$ gives for any choice of $w$, 
\begin{eqnarray}
\fl \Sfrac{1}{m_b+m_a}\, \log L_{m_b}^{[\dagger,1]} ({>}w,b) &+ 
  \Sfrac{1}{m_b+m_a}\, \log C_{m_a-u}^{(w)}(a) 
  \label{eqn19} \\
  &\leq \liminf_{t\to\infty}  \Sfrac{1}{(t+1)\,m_b+t\,m_a} 
  \log K^{(t)}(m_a,m_b,a,b,y)  
  =\Lambda_{inf}(m_a,m_b,a,b,y) . \nonumber
\end{eqnarray}

An upper bound is obtained by replacing the loops and walks in 
figure \ref{fig4} with adsorbing positive self-avoiding walks.  
This shows that, for $a,b\geq 0$ and $y\geq 0$,
\begin{eqnarray}
\fl \Lambda_{sup} (m_a,m_b,a,b,y)
= \limsup_{t\to\infty}  \Sfrac{1}{(t+1)\,m_b+t\,m_a} 
    & \log K^{(t)}(m_a,m_b,a,b,y) \nonumber \\
& \leq \Sfrac{1}{m_b+m_a}\, \log C_{m_b}^+(b) 
+ \Sfrac{1}{m_b+m_a}\, \log C_{m_a}^+ (a) ,
\label{eqn20}
\end{eqnarray}
where $C_n^+(a)$ is the partition function of adsorbing positive walks.
These bounds are valid in the cubic lattice.

\subsection{Limits in the cubic lattice}
\label{subsec:limits}

The limit $t\to\infty$ is an infinite comb with finite segments 
of length $m_b$ along the backbone, and teeth each of finite 
length $m_a$.  If $m_a\to\infty$ with $m_b =o(m_a)$ and diverging, 
then the limiting object will be a star with an infinite 
number of arms (this will be the \textit{star limit}), and, 
if instead, $m_b\to\infty$ with $m_a=o(m_b)$ and diverging, then 
the limit should be a \textit{self-avoiding walk limit}.  

One may also calculate limits of intermediate combs by putting 
$m_b=\lfloor\alpha\,n\rfloor$ and $m_a=\lfloor (1-\alpha)\,n\rfloor$ 
and then taking $n\to\infty$.   These limits are given in the 
following theorems.

Consider the star limit first.  

\begin{theo}[Star limit]
If $m_a\to\infty$ and $m_b=o(m_a)$ and divergent, the free energy is given by
\[ \fl \Lambda(a,b,y) 
= \lim_{m_a\to\infty} \Lambda_{inf} (m_a,o(m_a),a,b,y)
= \lim_{m_a\to\infty} \Lambda_{sup} (m_a,o(m_a),a,b,y) = \kappa(a) \]
for all $a,b\geq 0$ and $y > 0$.
\label{thm3star}

\begin{proof}
By equation \BRef{eqn11} a lower bound on the free energy in the limit as 
$m_a\to\infty$ (and $m_b=o(m_a)$) is, for all $y > 0$ and $a,b\geq 0$,

\[ \liminf_{m_a\to\infty} \Lambda_{inf}(m_a,o(m_a),a,b,y) \geq  
\liminf_{m_a\to\infty} \Lambda_{inf}(m_a,o(m_a),0,0,y) = \log \mu_3 = \kappa(1).\]
Notice that there is no dependence on $y$.

Similarly, by equation \BRef{eqn19} for $y > 0$ and $a,b\geq 0$,
\[\liminf_{m_a\to\infty} \Lambda_{inf}(m_a,o(m_a),a,b,y)
\geq \kappa^{(w)} (a) .\]
If $m_a\to\infty$, and $w \leq m_b=o(m_a)\to\infty$, it follows 
that $w$ can be increased without bound.  Since $\kappa^{(w)}(a)\to \kappa(a)$
by lemma \ref{T1}, this shows that for all $y>0$, and $a,b\geq 0$,
\[\liminf_{m_a\to\infty} \Lambda_{inf}(m_a,o(m_a),a,b,y) \geq \kappa(a).\]

Next, taking $m_a\to\infty$ with $m_b=o(m_a)\to\infty$, it follows from 
equation \BRef{eqn13} that for $y\geq 0$ and $a\geq 0$,
\[ \limsup_{m_a\to\infty} \Lambda_{sup} (m_a,o(m_a),1,1,y) \leq \kappa(a). \]
By equation \BRef{eqn20} 
\[ \limsup_{m_a\to\infty} \Lambda_{sup}(m_a,o(m_a),a,b,y) \leq \kappa(a) \]
for all $y\geq 0$ and $a,b\geq 0$.

Comparing the upper and lower bounds establishes that, for $y>0$ and $a,b\geq 0$,
\[ \lim_{m_a\to\infty} \Lambda_{inf}(m_a,o(m_a),a,b,y)
= \lim_{m_a\to\infty} \Lambda_{sup} (m_a,o(m_a),a,b,y) = \kappa(a) ,\]
by equations \BRef{eqn11}, \BRef{eqn13}, \BRef{eqn19} and \BRef{eqn20}.  
This completes the proof.
\end{proof}
\end{theo}

The star limit in theorem \ref{thm3star} shows that there is a critical plane
$a=a_c$ in the $aby$-phase diagram where the star adsorbs. There is no 
ballistic phase for any finite $y\geq 0$, so that the star cannot be pulled 
from the adsorbing plane. The node of the star is the limit of the
backbone of the comb, and since the backbone is both grafted to the
adsorbing surface, and pulled by the vertical force, it cannot be pulled
from the adsorbing surface. 

Next, consider the limits when $m_b=\lfloor \alpha\,n\rfloor$ 
and $m_a=\lfloor (1-\alpha)\,n\rfloor$ for an $\alpha\in(0,1)$.  Then
taking $n\to\infty$ also take $m_b$ and $m_a$ to infinity.  By 
equations \BRef{eqn11} and \BRef{eqn19},
\begin{equation}
\liminf_{n\to\infty} \Lambda_{inf}(\lfloor \alpha n \rfloor,
\lfloor (1-\alpha)n\rfloor,a,b,y) \geq
\cases{
\alpha\,\kappa(b) + (1-\alpha)\,\kappa^{(w)} (a) \\
\alpha\,\lambda(y) + (1-\alpha)\,\log \mu^{(w)} } 
\end{equation}
for all $a,b,y\geq 0$.  If $y\geq 1$, then by lemma \ref{lemma2a} 
the most popular width of the loops in figure \ref{fig4} 
increases without bound as $m_b\to\infty$ in both the adsorbed 
and desorbed phases.  Thus, the limit $w\to\infty$ can be taken 
on the right hand side in the above, giving
\begin{equation}
\liminf_{n\to\infty} \Lambda_{inf}(\lfloor \alpha n \rfloor,
\lfloor (1-\alpha)n\rfloor,a,b,y) \geq
\cases{
\alpha\,\kappa(b) + (1-\alpha)\,\kappa (a) ;\\
\alpha\,\lambda(y) + (1-\alpha)\,\log \mu_3 ,}
\label{eqn27}
\end{equation}
for $y\geq 1$ and $a,b\geq 0$.

More generally, notice that
\[ K^{(t)}(\lfloor\alpha n\rfloor,\lfloor (1-\alpha)n\rfloor,a,b,y)
\geq K^{(t)}(\lfloor\alpha n\rfloor,\lfloor (1-\alpha)n\rfloor,0,0,0)\]
and that $K^{(t)}(\lfloor\alpha n\rfloor,\lfloor (1-\alpha)n\rfloor,0,0,0)$
is the number of combs from the origin to their last vertex
in the plane $x_3=0$, and with no visits by teeth or the backbone
to the plane $x_3=0$ otherwise.  Then it follows, by using
the methods of reference \cite{Soteros}, that
\begin{eqnarray}
 & \liminf_{n\to\infty} 
  \Sfrac{1}{(t+1)\lfloor\alpha n\rfloor + t\,\lfloor (1-\alpha)n\rfloor} 
 \log K^{(t)}(\lfloor\alpha n\rfloor,\lfloor (1-\alpha)n\rfloor,a,b,y) \nonumber \\
& \geq \liminf_{n\to\infty} 
  \Sfrac{1}{(t+1)\lfloor\alpha n\rfloor + t\,\lfloor (1-\alpha)n\rfloor}  
 \log K^{(t)}(\lfloor\alpha n\rfloor,\lfloor (1-\alpha)n\rfloor,0,0,0) \nonumber \\
& \geq \log \mu_3 .
\end{eqnarray}
This shows, that for every $\epsilon>0$ there is a $N_\epsilon$
such that for all $n\geq N_\epsilon$ and $a,b,y\geq 0$, 
\begin{equation}
K^{(t)}(\lfloor\alpha n\rfloor,\lfloor (1-\alpha)n\rfloor,a,b,y)
\geq \left( (t+1)\lfloor\alpha n\rfloor + t\,\lfloor (1-\alpha)n\rfloor\right
(\log \mu_3 - \epsilon) .
\end{equation}
Thus, by the definition of 
$\Lambda_{inf} (\lfloor\alpha n\rfloor,\lfloor (1-\alpha)n\rfloor,a,b,y)$
if follows that, for all $n\geq N_\epsilon$ and $a,b,y\geq 0$,
\begin{eqnarray}
&\Lambda_{inf} (\lfloor\alpha n\rfloor,\lfloor (1-\alpha)n\rfloor,a,b,y)
 \nonumber \\
&= \liminf_{t\to\infty} 
  \Sfrac{1}{(t+1)\lfloor\alpha n\rfloor + t\,\lfloor (1-\alpha)n\rfloor} 
 \log K^{(t)}(\lfloor\alpha n\rfloor,\lfloor (1-\alpha)n\rfloor,a,b,y)
  \nonumber \\
&\geq \log \mu_3 - \epsilon .
\end{eqnarray}
Since $\epsilon>0$ is arbitrary, it follows that for all $a,b,y\geq 0$,
\begin{equation}
\liminf_{n\to\infty} 
\Lambda_{inf} (\lfloor\alpha n\rfloor,\lfloor (1-\alpha)n\rfloor,a,b,y)
\geq \log \mu_3 .
\end{equation}
Comparing this to the right hand side of equation \BRef{eqn27} 
shows that this is a lower bound for all $a,b,y\geq 0$, and hence 
equation \BRef{eqn27} is valid also for all $y\geq 0$.

\begin{theo}
The limit
\begin{eqnarray*}
\Lambda_\alpha (a,b,y)
&=\lim_{n\to\infty} 
\Lambda_{sup} (\lfloor \alpha n\rfloor,\lfloor(1-\alpha)n\rfloor,a,b,y)
\nonumber \\
&=\max\{\alpha\,\lambda(y) + (1-\alpha)\,
\log \mu,\alpha\,\kappa(b) + (1-\alpha)\,\kappa(a) \} 
\end{eqnarray*}
exists.
\label{T5}

\begin{proof}
Consider first the case that $a,b\leq 1$.  By equation \BRef{eqn13},
\[ \limsup_{n\to\infty} 
\Lambda_{sup} (\lfloor \alpha n\rfloor,\lfloor(1-\alpha)n\rfloor,a,b,y) \leq
\alpha\,\lambda(y) + (1-\alpha)\,\log \mu . \]
On the other hand, if $a>a_c$, or $b>b_c$, or both, then by equation \BRef{eqn20},
\[ \limsup_{n\to\infty} 
\Lambda_{sup} (\lfloor \alpha n\rfloor,\lfloor(1-\alpha)n\rfloor,a,b,y) \leq
\alpha\,\kappa(b) + (1-\alpha)\,\kappa(a) . \]
These upper bounds coincide with the lower bounds in equation \BRef{eqn27}.
Since these bounds are monotonic non-decreasing in each of $\{a,b,y\}$,
and since $\kappa(a)=\kappa(b)=\log\mu$ for all $a\leq a_c$ and $b\leq b_c$
the result is that
\begin{eqnarray*}
\fl
\liminf_{n\to\infty} 
\Lambda_{inf} (\lfloor \alpha n\rfloor,\lfloor(1-\alpha)n\rfloor,a,b,y)
&= \limsup_{n\to\infty} 
\Lambda_{sup} (\lfloor \alpha n\rfloor,\lfloor(1-\alpha)n\rfloor,a,b,y)
\nonumber \\
&= \max\{\alpha\,\lambda(y) + (1-\alpha)\,
\log \mu,\alpha\,\kappa(b) + (1-\alpha)\,\kappa(a) \} . 
\end{eqnarray*}
This completes the proof.     
\end{proof}
\end{theo}

Notice that by taking $\alpha \to 0^+$ the star limit in theorem \ref{thm3star} is
found.

The self-avoiding walk limit in the cubic lattice is encountered when
$m_b\to\infty$ and $m_a=o(m_b)$ and divergent.  By taking $\alpha\to 1^-$
in theorem \ref{T5}, it follows that 

\begin{theo}[Self-avoiding walk limit]
If $m_b\to\infty$ and $m_a=o(m_b)$ and divergent, the free energy is given by
\begin{eqnarray*}  
\Lambda(a,b,y) 
&= \lim_{m_b\to\infty} \Lambda_{inf} (o(m_b),m_b,a,b,y) \\
&= \lim_{m_b\to\infty} \Lambda_{sup} (o(m_b),m_b,a,b,y) 
= \max\{\kappa(b),\lambda(y)\} = \psi(b,y).
\end{eqnarray*}
for all $a,b\geq 0$ and $y\geq 0$. \hfill \qed
\label{thm3walk}
\end{theo}

\section{Discussion}
\label{sec:discussion}

We have investigated a self-avoiding walk model of the adsorption 
of comb copolymers at a surface.  These are interesting because 
of their use as steric stabilizers of dispersions \cite{Xie}.  The 
teeth of the comb are chemically different from the backbone of 
the comb and can interact differently with the surface.  In addition, 
we have considered the effect of a force pulling the adsorbed comb off 
the surface.  We have concentrated on the case where one end vertex 
of the backbone is fixed in the surface and where the force is 
applied at the other end of the backbone.

We have considered several cases: (i) a uniform comb where the 
lengths of the teeth and the backbone branches (or segments) are 
all equal and where the number of teeth ($t$) is fixed, (ii) a 
comb with $t$ teeth where the teeth are short compared to the 
backbone branches, (iii) a comb with $t$ teeth where the backbone 
branches are short compared to the teeth, and (iv) the case where 
the number of teeth goes to infinity.  In each case we have 
determined the free energy rigorously and investigated the forms 
of the phase diagrams.  In particular, for the uniform case 
(where the lengths of the teeth and the backbone branches are 
equal and where the number of teeth is fixed) we have established 
the form of the phase diagram by taking various slices through the 
three dimensional diagram.  When the teeth do not adsorb, the 
form of the (two dimensional) diagram is similar to that of 
self-avoiding walks adsorbing in a surface and being desorbed 
by a force \cite{Rensburg2013}.  When both the teeth and the 
backbone adsorb, there are strong similarities to the phase 
diagram for copolymeric stars \cite{Rensburg2022}.  

Comparing the infinite number of teeth case (iv) to the other 
cases considered, our results establish that some limits can be 
interchanged.  For case (ii), when the teeth are short compared 
to the backbone branches, then the walk limit is obtained whether 
one first lets the number of teeth ($t$) go to infinity and then lets 
the backbone length go to infinity (as in section \ref{subsec:limits}, 
theorem \ref{thm3walk}) or if the limits are in the reverse order 
(see section \ref{sec:nonuniform}, theorem \ref{theo:teethslow}).  
Similarly, for case (iii), when the the backbone branches are short 
compared to the teeth, no matter the order of the limits, the star 
limit is obtained (see theorems \ref{thm3star} and \ref{theo:sec4star}).
For the uniform case (i), comparing theorems \ref{T5} 
($\alpha =1/2$)  and \ref{theo:uniformFE}, no matter the order 
of the limits ($t \to \infty$,  length of the branches goes to infinity)  
the free energy is the same. The resulting comb free energy is 
either equal to that of the adsorbed case ($y=1$) or that of 
the pulled/non-adsorbed case ($a=b=1$).     

Although we have concentrated on the case of the simple 
cubic lattice our results can be extended to the $d$-dimensional 
hypercubic lattice for all $d \ge 3$, but our methods do not 
extend to the $d=2$ case.  The two dimensional case requires further 
work.  In three dimensions, our methods could be extended to apply 
to other lattices such as the body centred and face centred cubic 
lattices, although we have not worked out the details.

The model can be extended in other ways, for instance to a brush 
copolymer where we have a backbone with $t$ vertices of degree $k$ 
and with $k-2$ side chains (the analogue of teeth of a comb) at 
each of these $t$ vertices.  It would also be interesting to 
consider cases where the force is applied at vertices other than the 
terminal vertex of degree 1.

\section*{Acknowledgement}
EJJvR and CES acknowledge the Natural Sciences and Engineering 
Research Council of Canada (NSERC) [funding reference numbers: 
RGPIN-2019-06303; RGPIN-2020-06339].

\end{document}